\documentclass[useAMS,usenatbib]{mn2e}
\usepackage{graphicx,url,longtable}

\newcommand{\aap}{A\&A}
\newcommand{\aj}{AJ}
\newcommand{\apj}{ApJ}
\newcommand{\apjl}{ApJL}
\newcommand{\apjs}{ApJS}
\newcommand{\araa}{ARA\&A}
\newcommand{\mnras}{MNRAS}
\newcommand{\iaucirc}{IAUC}

\newcommand{\kmsmpc}{\>{\rm km}\,{\rm s}^{-1}\,{\rm Mpc}^{-1}}
\newcommand{\kms}{\>{\rm km}\,{\rm s}^{-1}}
\newcommand{\flux}{\rm {erg\,cm^{-2}\,s^{-1}}}

\newcommand{\kev}{\rm{keV}}
\newcommand{\beq}{\begin{equation}}
\newcommand{\eeq}{\end{equation}}
\def\log{\,{\rm log}\,}

\begin{document}
\title{Radio loud AGN and the $L_X - \sigma$ relation of galaxy groups and clusters}
\author[Shen et al.]
{Shiyin Shen$^{1,2,3}$, Guinevere Kauffmann$^{2}$, Anja von der Linden$^{2}$,
Simon D.M. White$^{2}$, \and P.N. Best$^{4}$
\thanks {E-mail: ssy@shao.ac.cn}
 \\
$^1$ Shanghai Astronomical Observatory, Chinese Academy of Sciences, Shanghai 200030, China\\
$^2$ Max-Planck-Institut f\"ur Astrophysik, Karl Schwarzschild Str. 1, Postfach 1317, 85741 Garching, Germany\\
$^3$ Joint Institute for Galaxy and Cosmology of the Shanghai Astronomical
Observatory and the University of Science and Technology of China \\
$^4$ SUPA, Institute for Astronomy, Royal Observatory Edinburgh, Blackford
Hill, Edinburgh, EH9 3HJ }

\maketitle

\begin{abstract}

We use the ROSAT All-Sky Survey to study the X-ray properties of a sample of
625 groups and clusters of galaxies selected from the Sloan Digital Sky Survey.
We stack clusters with similar velocity dispersions and investigate whether
their average X-ray luminosities and surface brightness profiles vary with the
radio activity level of their central galaxies. We find that at a given value
of $\sigma$, clusters with a central radio AGN have more concentrated X-ray
surface brightness profiles, larger central galaxy masses, and higher X-ray
luminosities than clusters with radio-quiet central galaxies. The enhancement
in X-ray luminosity is more than a factor of two, is detected with better than
6$\sigma$ significance, and cannot be explained by X-ray emission from the
radio AGN itself. This difference is largely due to a subpopulation of
radio-quiet, high velocity dispersion clusters with low mass central galaxies.
These clusters are underluminous at X-ray wavelengths when compared to
otherwise similar clusters where the central galaxy is radio-loud, more
massive, or both.

\end{abstract}

\section{introduction}

There is increasing evidence that the majority of radio-loud AGN at low
redshift may be triggered by the accretion of hot gas. Using a sample of 625
nearby groups and clusters, \cite{Best07} showed that the galaxies located
closest to the centres of the clusters are more likely to host a radio-loud AGN
than other galaxies of similar stellar mass.  \cite{Allen06} analyzed
\emph{Chandra} X-ray images of nine nearby X-ray luminous elliptical galaxies
and showed that the jet power is tightly correlated with the Bondi accretion
rate onto the central black hole as estimated from the observed gas temperature
and density profiles.

It is also now recognized that the `over-cooling problem' at the centre of many
galaxy clusters could be solved if radio-loud AGN can heat surrounding gas.
Direct evidence of AGN heating came with the discovery of X-ray cavities in the
hot intracluster medium of a number of clusters and groups
\citep{Boehringer93,Churazov00,Fabian00,McNamara00,Blanton03,Gitti07}. Some of
these cavities coincide with extended lobes of radio emission produced by an
AGN in the central cluster galaxy\footnote{Some cavities do not have associated
radio emission above current detection limits [e.g. \cite{Ettori02},
\cite{Clarke05} and \cite{Jetha08}] but these hot bubbles may have been
inflated by a previous cycle of nuclear activity.}. The observations suggest
that relativistic radio plasma displaces the X-ray emitting gas, generating
turbulence and wave activity which heat the intracluster medium (ICM).

AGN feedback may also explain why low-temperature clusters are less X-ray
luminous than predicted by a homologous scaling of the properties of hotter
and more massive systems \citep{Nath02,Best07}. The energy available from the
central AGN depends primarily on black hole accretion rate. In low-mass
groups, it can be comparable to the total gravitational binding energy of the
X-ray emitting gas, whereas high-mass clusters will be less strongly affected.
As a result, relations such as $L_X-T_X$ (X-ray luminosity vs temperature) and
$L_X-\sigma$ (X-ray luminosity vs velocity dispersion) may be modified by
radio-source heating. At smaller scales, AGN feedback may play an important
role in the formation and evolution of galaxies. Recent work has demonstrated
that it can plausibly explain the exponential cutoff at the bright end of the
galaxy luminosity function, as well as the `down-sizing' of galactic
star-formation activity at recent cosmic epochs \citep{Croton06, Bower06,
Kang06}.

It may be that hot gas accretion, AGN triggering, reheating of ambient gas,
and suppression of AGN activity occur in a cycle. Accretion onto the central
black hole may cause an outburst which removes the fuel supply for
further radio activity.  Without AGN feedback, the ICM reverts to the
state that triggered the outburst \citep{Churazov05}. Although this picture is
attractive, it remains to be verified in detail. We do not know the necessary
conditions to trigger or to quench radio activity. We also do not understand
the extent to which the global X-ray properties of the gas are modified by AGN
feedback\citep[e.g.][]{Rizza00,Omma04,Nusser08}. Only a few nearby clusters
have deep enough X-ray images to reveal low-density cavities and permit the
energetics of the gas to be studied in a spatially resolved fashion. If we
wish to study how the global state of the X-ray gas is linked to AGN activity
in the central galaxy, we are forced to adopt a more statistical approach.

It has long been known that the radio activity of cluster central galaxies is
linked with the cooling flow or ``cool core'' phenomenon
\citep{Burns90,Fabian94a}. This is the fact that many but not all massive
clusters show a strong central peak in X-ray surface brightness, almost always
coincident with the brightest cluster galaxy (BCG), within which the directly
inferred cooling time is less than the Hubble time. The fraction of cool-core
clusters with a radio-loud BCG is much higher than the fraction in the rest of
the population. Furthermore, because the cores add significantly to cluster
luminosity and reduce the emission-weighted temperature, cool-core clusters
fall systematically to one side of many of the X-ray scaling relations for
clusters, e.g.  the $L_X-T_X$ and $L_X-\sigma$ relations\citep{Fabian94b,
OHara06, Chen07}. Thus, in clusters of given velocity dispersion (or mass),
cool cores tend to be associated with enhanced X-ray luminosity, with massive
central BCGs, and with radio-loud BCGs. Given that the probability of radio
activity increases with BCG mass\citep{Best05}, that both BCG mass and cluster
X-ray luminosity increase with cluster mass\citep{Edge91}, and that central
black hole mass correlates tightly with galaxy mass\citep{Tremaine02}, it seems
clear that radio activity is tightly related to both black hole mass and the
presence of a cooling hot atmosphere which can provide fuel. The exact causal
relation between these phenomena remains unclear, however, and recent
observational studies of lower mass systems appear, at least superficially, in
conflict with the trends established for relatively massive clusters [e.g.
compare \cite{Croston05} and \cite{Jetha07} with the above references].

In this paper, we provide improved statistics for the relationship between
cluster properties, both X-ray and optical, and the radio activity of their
BCGs. We use a low redshift ($z<0.1$) sample of 625 groups and clusters with
carefully controlled BCG identifications. These were selected by
\cite{Linden07} from the Sloan Digital Sky Survey\citep[SDSS,][]{York00}.
Radio properties of these BCGs were determined following \cite{Best05}, who
cross-matched the galaxies with the National Radio Astronomy Observatory Very
Large Array Sky Survey \citep[NVSS,][]{Condon98} and the Faint Images of the
Radio Sky at Twenty Centimeters Survey \citep[FIRST,][]{Becker95}. By
combining these two radio surveys, the identification of radio galaxies is
both reasonably complete ($\sim95$ per cent) and highly reliable ($\sim99$ per
cent). Radio-loud AGN are then separated from star-forming galaxies with
detectable radio emission on the basis of 4000\AA\ break
strength\citep{Best05}.

The sky coverage of this cluster sample is $\sim4000$ square degrees.  The
only X-ray survey with enough sky coverage to provide a reasonable match is
the ROSAT All-Sky survey(RASS). Because the RASS is quite shallow, relatively
few individual groups and clusters have sufficient X-ray flux for unambiguous
detection, particularly at low velocity dispersion. The mean flux of such
objects {\it can} be detected, however, by stacking their X-ray images
\citep[e.g.][]{Bartelmann03, Shen06, Dai07, Rykoff08}. Such stacking avoids
selection biases that can occur if one analyses samples in which only the most
X-ray luminous systems are detected. The limited resolution of RASS maps and
our stacking strategy, make it difficult to study the structure of the ICM in
detail, but we will see that concentration differences, reflecting the presence
of cool cores, are nevertheless detectable.

Our paper is organized as follows. In Section 2, we introduce our sample of
groups and clusters and describe their radio properties. In Section 3, we
describe our X-ray detection techniques, both our method for detecting
individual groups and clusters and our stacking technique. In section 4, we
study and compare the X-ray properties of clusters, emphasising how the
$L_X-\sigma$ relation and the surface brightness profiles of clusters vary
with the radio properties of their BCGs. We discuss the contribution of the
radio AGN to the total X-ray emission in Section 5.1, while in Section 5.2, we
discuss how the $L_X-\sigma$ relation of clusters depends on the stellar
properties of their BCGs. We present our conclusions in Section 6.

\section{sample}
We use the sample of groups and clusters of galaxies described in
\cite{Linden07}, which is drawn from the C4 cluster catalogue of the SDSS Data
Release 3 \citep{Miller05}. The clusters lie in the redshift range $0.02 \leq
z \leq 0.1$. Von der Linden et al. developed improved algorithms for
identifying the brightest galaxy (the BCG) and for measuring the velocity
dispersion in each group or cluster. The velocity dispersion algorithm is
designed to limit the effects of neighbouring groups and clusters. Clusters
and groups with very few galaxies are also discarded. The velocity dispersion
$\sigma$ is measured within the virial radius $R_{200}$. The radio properties
of the BCGs are taken from the catalogue of \cite{Best05}, which has been
updated to the SDSS data release 4.

The final sample includes 625 groups and clusters, with velocity dispersion
spanning the range from $\sim 200\,\kms$ to over 1000 $\kms$ (see von der
Linden et al. 2007 for more details). 134 out of the 625 BCGs have radio
fluxes larger than 5 mJy and are identified as radio-loud AGN. Five BCGs have
radio fluxes which exceed 5 mJy but are clearly a result of star formation
activity.  There are 433 BCGs without a radio source brighter than 5mJy. The
radio properties of the remainder are unknown because they lie outside the
area covered by FIRST. Henceforth, we will refer to a group or cluster as
`radio-loud' if its BCG has been identified as a radio-loud AGN, and `radio
quiet' if it is known to contain no radio source brighter than 5mJy. There are
134 radio-loud and 433 radio-quiet clusters in our sample. The fraction of
radio-loud clusters in our sample is smaller than the fraction of clusters
usually quoted as having cooling cores\citep[$\sim40\%$ according
to][]{Peres98}. This may well be a selection effect since X-ray selected
cluster samples are biased towards massive and X-ray luminous systems.

\begin{figure}
  \includegraphics[width=84mm]{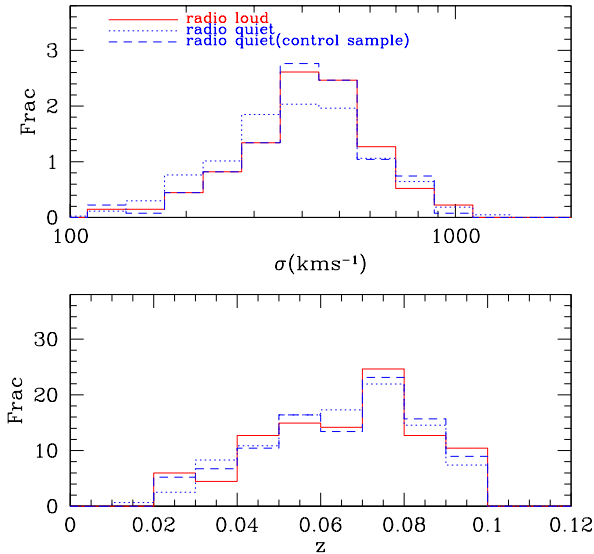}\\
\caption{Velocity dispersion and redshift distributions for all radio-loud
clusters (solid histograms), for all radio-quiet clusters (dotted histograms)
and for our matched control sample of radio-quiet clusters (dashed histograms).}
  \label{RLRQ}
\end{figure}

In figure \ref{RLRQ} we show histograms of velocity dispersion $\sigma$ and
redshift $z$ for radio-loud and radio-quiet clusters.  The two redshift
distributions are very similar. The radio-loud clusters have slightly higher
velocity dispersions than the radio-quiet objects. The median $\sigma$ of
radio-loud clusters is 428 $\kms$, with 16 and 84 percentiles at 277 and 595
$\kms$ respectively. For radio-quiet clusters, the corresponding $\sigma$
values are 392, 252 and 583 $\kms$. The figure also shows $\sigma$ and $z$
distributions for a control sample of radio-quiet clusters (dashed line). This
was constructed by choosing the radio-quiet cluster closest in $\sigma$ and $z$
to each radio-loud cluster (see Section \ref{Sta}). This matching procedure
was introduced in order to minimize possible observationally induced biases
when comparing the two samples.

\section{X-ray detection}
We use the RASS to study the X-ray properties of our sample of groups and
clusters. The RASS mapped the sky in the soft X-ray band (0.1$-$2.4keV) with
exposure time varying between 400 and 40,000s, depending on direction on
the sky. The resolution of the RASS images is 45 arcsec. Two source catalogues
were generated based on RASS images through a maximum-likelihood (ML) search
algorithm: the bright source catalogue \citep{Voges99} and the faint source
catalogue \citep{Voges00}. However, these source catalogues are not optimized
for extended sources. An extended source is likely to be deblended into several
pieces by the ML search algorithm.

There are carefully selected  samples of X-ray clusters constructed from the
RASS data, e.g. the Northern ROSAT All-Sky Galaxy Cluster Survey
\citep[NORAS,][]{Boehringer00} and the ROSAT-ESO Flux-Limited X-Ray Galaxy
Cluster Survey \citep[REFLEX,][]{Boehringer04}). The REFLEX clusters, which
have a completeness of $\geq90\%$ at the flux limit of $3\times10^{-12}\flux$,
are mainly located in the southern sky ($\delta\leq2.5$) and have little
overlap with our SDSS clusters. For the NORAS clusters, the completeness is
estimated to be $\sim 50\%$ at an X-ray flux of $3\times10^{-12}\flux$
\citep{Boehringer04}.

Our sample of groups and clusters has been selected from an optical galaxy
catalogue. The centre of the cluster (taken to be the position of the BCG) and
the virial radius of each cluster have already been determined. We use this
extra information when studying their X-ray properties. We first check whether
the position of the BCG of each cluster is consistent with a peak in X-ray
emission (Section \ref{Xcent}). We then optimize the growth curve analysis
algorithm of \cite{Boehringer04} to detect the X-ray properties of our
clusters individually (Section \ref{Ind_det}). Finally, we develop a stacking
algorithm to measure the mean X-ray properties of clusters as a function of
their optical properties, independent of the detection limit of the RASS
(Section \ref{Sta_det}).

\subsection{Finding the X-ray centre of the clusters}\label{Xcent}
Before the X-ray luminosity of a cluster can be measured, the position of its
centre must be determined. In the optical, the position of the BCG is usually
taken to define the cluster centre. However, this position may be offset from
the peak of the X-ray emission \cite[e.g.][]{Dai07, Koester07}. In order
understand whether such offsets are generic for clusters in our sample, we have
developed an algorithm to determine the X-ray centre of a cluster.

First, the ML search algorithm used to generate the ROSAT source catalogue is
applied to the RASS images \citep{Voges99}. Sources with detection likelihood
$L>7$ in the $0.5-2.0$ keV band are retained. Although an extended source is
frequently de-blended into several pieces by this algorithm, the central peak
is always identified. We match the position of each BCG with the ML detections
using a tolerance of 5 arcmin in radius. X-ray sources found within this radius
are considered as candidate X-ray centres. We find that 210 clusters have X-ray
sources within a 5 arcmin radius and some clusters have several candidate X-ray
centres. We then exclude candidates which are clearly not associated with the
clusters by eye. If a point source is detected within the 5 arcmin search
radius but significantly offset from the BCG, we assume that the source is a
contaminant and is not part of the cluster. We show such an example in Fig. B1
of Appendix B. For clusters that still have more than one candidate X-ray
centre, we pick the closest one. Most of the sources we identified as the X-ray
centres of  clusters have extension likelihood greater than zero. This confirms
that the X-ray emissions we identified are from the extended ICM rather than
from point-like AGN. As we will show in Section \ref{AGNcont}, the X-ray
emission from radio AGN in the soft X-ray bands is expected to be much weaker
than the X-ray emission we measure. Our final sample of clusters with
identified X-ray centres contains 157 objects.

\begin{figure}
  \includegraphics[width=84mm]{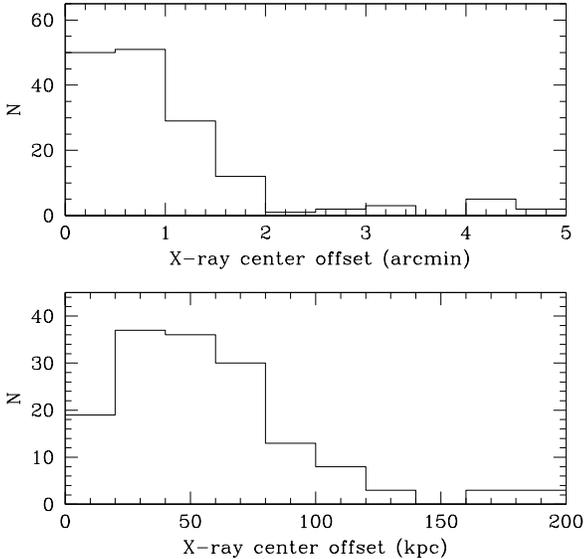}\\
\caption{A histogram of the offset between BCG and X-ray centre. The upper
panel shows the offset in units of arcmin, while the lower panel shows the
offset in units of kpc.}
  \label{offset}
\end{figure}

In Figure~\ref{offset} we plot the offset between X-ray centre and optical
centre (i.e. the BCG position) for these 157 clusters. The upper panel shows
the offset in units of arcmin, while the lower panel shows the corresponding
projected physical distance at the redshift of BCG. As can be seen, the X-ray
centre is consistent with the BCG position in most cases. The typical offset is
less than 2 armin ($\sim$ 3 RASS pixels).  This corresponds to a physical
distance of around 50 kpc and so is consistent with the results of
\cite{Katayama03}.

Since the BCG positions are, in general, consistent with the X-ray centres, we
will assume that the BCGs mark the centres of those clusters which are too
faint to determine an X-ray centre directly with our algorithm.

\subsection{Measuring the X-ray luminosity}\label{Ind_det}

After the centre of each cluster has been  determined, we use a growth curve
analysis to count the number of photon events as function of  radius and
so to evaluate the cluster count rate.

The background of each cluster is estimated from a centred annulus with inner
radius $R_{200}$ and width 6 arcmin. To exclude contamination of this
background estimate by discrete sources, we mask out all ML-detected ($L>7$)
sources inside this annulus. The surface brightness of the background is then
calculated using the formula
\begin{equation}
 \mu = \frac{\sum_{i=1}^N \frac{1}{t_i}}{S},
\end{equation}
where the sum is over the photon events, $t_i$ is the effective exposure time
at each photon position, and $S$ is the effective area of the annulus after
the contaminating sources are masked out.

The count rates within $R_{200}$ may also be contaminated by discrete X-ray
sources unassociated with intracluster X-ray emission. As above, we therefore
mask out all the ML-detected ($L>7$) {\it point} sources that are located more
than 2.25 arcmin (3 image pixels) from the X-ray centre\footnote{Here, we
assume that all X-ray photons inside the 2.25 armin circle are emitted from
the ICM. The probability that a foreground or background source is located
inside this small circle by chance is close to zero. The X-ray emission from
the central AGN itself is expected to be negligible compared with that from
the ICM(see Section \ref{AGNcont}).}. As already mentioned, an extended source
is often detected multiple times by the ML algorithm. We therefore visually
inspect all the ML sources and determine whether they are contaminants or part
of the X-ray emission from the cluster. We show an example of this process in
Fig.~B2 of Appendix B.

The cumulative source count rate as a function of radius is calculated by
integrating the source counts in concentric rings and subtracting the
background contribution. We integrate the source count rate to the X-ray
extension radius $R_X$. We determine this radius in two different ways, as
described in \cite{Boehringer00}. The first chooses the radius where the
increase in source signal is less than the 1$\sigma$ uncertainty in the count
rate. The other is the plateau-fitting method, where the slope of the plateau
(in units of count rate per arcmin) is less than 1 percent. These two methods
give consistent results for clusters where the count rate profile can be
determined with high S/N. For low S/N clusters, we choose the better
determination of $R_X$ by visually checking the cumulative count rate profile.
A source is said to have a significant X-ray detection if the counts within
$R_X$ are three times larger than the Poisson fluctuation in the photon counts.
We detect 159 galaxy clusters individually according to this criterion.

To convert the count rates into X-ray fluxes, we assume that the X-ray emission
has a thermal spectrum with temperature $T$, that the cluster gas has metal
abundance equal to a third of the solar value \citep{Raymond77} and that
interstellar absorption can be inferred from the Galactic hydrogen column
density \citep{Dickey90}. The gas temperature $T$ is assumed to follow the
empirical scaling relation measured by \cite{White97},

\begin{equation}
T=\left(\frac{\sigma}{403\kms}\right)^2.
\end{equation}

One might question whether this assumption is robust. Most studies of the
$T-\sigma$ relation find that it does not depart significantly from the virial
theorem expectation ($T\propto \sigma^2$), even for low-mass galaxy groups
\citep[e.g.][]{Girardi96, Wu99, Mulchaey00, Xue00}. We further note that a
variation in temperature of 50 percent makes less than a 5 percent difference
to the flux estimate for most of our clusters.\footnote{If the gas in
radio-loud groups is hotter than in radio-quiet groups as suggested, for
example, by \cite{Croston05}, we will underestimat the X-ray luminosity of
radio-loud clusters and the conclusion we reach below about the difference in
X-ray luminosity between radio-loud and radio-quiet clusters (Section
\ref{LSstack}) will be reinforced.} Throughout this paper we adopt a
concordance $\Lambda$CDM cosmology with $H_0=70\kmsmpc$, $\Omega_0=0.3$, and
$\Omega_\Lambda=0.7$.

For some clusters, the total X-ray flux will exceed the flux measured within the
extension radius $R_X$. To estimate the flux that is missed outside $R_X$, we
adopt a $\beta$-model surface brightness distribution $I_X(R)$ with
$\beta=2/3$,
\begin{equation}
I_X(R)=I_0(1+\frac{R^2}{R_c^{2}})^{-3\beta+0.5}, \label{betaprof}
\end{equation}
where $I_0$ is the central surface brightness and $R_c$ is the core radius and
is assumed to be proportional to $R_{200}$, with $R_c=0.14R_{200}$ for
radio-loud clusters and $R_c=0.18R_{200}$ for radio-quiet
clusters.\footnote{Here, the different $R_c$ values for radio-loud and
radio-quiet clusters are based on the results from stacked images shown in
Section \ref{SBprof}. This difference introduces a typical change of $L_X$ less
than 5 percent. We would come to exactly the same conclusions if we adopted the
same $R_c$ relation for all clusters, regardless of their radio properties.}  The
extension correction factor $f_E$ is then defined as
\begin{equation}
f_E=\frac{\int_0^{R_{200}}I_X(R) R dR}{\int_0^{R_X}I_X(R) R dR}.
\label{fe}
\end{equation}
The exact values of $R_c$ and $\beta$ are far from certain for each individual
cluster. As a result, the X-ray luminosities of clusters with large correction
factors $f_E$ have larger errors. To account for this effect, we assume that
the correction factor $f_E$ has an uncertainty of 0.1$f_E$. We note that for a
few clusters with sufficiently high S/N, we can apply a $\beta$ model fit and
derive the extension correction factor $f_E$. We find answers that are
consistent with the estimates using Equ. (\ref{fe}) with an uncertainty of
$\sim 10$\%.  The error on our estimate of $L_X$ for each cluster thus has two
terms: the error of the flux estimation inside $R_X$ and the error on the
correction factor $f_E$.

The X-ray properties of the 159 clusters that are individually detected in RASS
are listed in Appendix A.

\subsection{Stacking Analysis}\label{Sta_det}

The clusters that are individually detected in the RASS are biased to the
nearest and most X-ray luminous systems at each velocity dispersion. We can
obtain an unbiased estimate of the average X-ray luminosity of our clusters by
stacking objects with similar optical properties. We describe our stacking
algorithm below.

We divide the clusters into bins of velocity dispersion with a width of $\sim
100\kms$. Within each velocity dispersion bin, the angular size of the clusters
varies substantially because of the spread in redshift. (Recall that $R_{200}$
is proportional to $\sigma$.) We centre each cluster image on the X-ray centre
if this has been identified (see Section \ref{Xcent}) and on the BCG otherwise.
Before stacking the images, we scale them all to the same size in units of
$R_{200}$ and mask contaminating sources in the same way as in our analysis of
individual clusters. The background is estimated as the average flux inside an
annulus with inner radius $R_{200}$ and outer radius $1.3R_{200}$. The
remaining steps in the detection algorithm (determining the extension radius
$R_X$ and count rate) parallel those used for individual clusters. A stacked
image is said to have a robust X-ray detection if the number of source photons
within $R_X$ is also three times larger than the Poisson fluctuation in the
photon counts.

For a stack of $N$ clusters with individual X-ray luminosities $L_{X,i}$,
redshifts $z_i$, average Galactic hydrogen columns $N_{H,i}$ and average RASS
effective exposure times $t_i$, the number of source photons that should be
detected in the RASS, $N_s$,  is
\begin{equation}
\label{stack} N_s=\sum_{i=1}^N{L_{X,i}~g(N_{H,i}, z_i)t_i}\,,
\end{equation}
where $g(N_{H,i}, z_i)$ is a function which converts the bolometric X-ray
luminosity  to observed count rate. Since the X-ray luminosities of the sources
in a stack are similar, the weighted average X-ray luminosity of
the stack can be defined as

\begin{equation}\label{Lstack}
L_{X, S}=\frac{N_s}{\sum_{i=1}^N{g(N_{H,i}, z_i)t_i}}.
\end{equation}

We fit the surface brightness profile of each stack with a $\beta$ model to
make the extension correction, and we correct the X-ray luminosity to the
value expected within $R_{200}$ (see Section \ref{SBprof}).

\section{X-ray properties of the  clusters}
In this section, we investigate how the X-ray properties of the clusters
depend on the radio properties of their central BCGs. We focus on the
comparison of the $L_X-\sigma$ relations of radio-loud and radio-quiet
clusters. We first show results for clusters that were detected individually
in the X-ray images (Section \ref{Ind}). We then present results for stacked
cluster samples (Section \ref{Sta}).

\subsection{Results for individual X-ray detections}\label{Ind}

 \subsubsection{Detected fraction}\label{frac}

\begin{figure}
  \includegraphics[width=84mm]{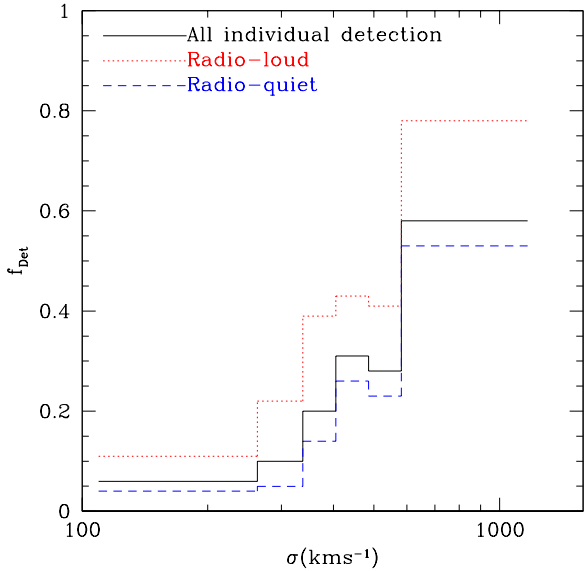}\\
\caption{The fraction of clusters with individual X-ray detections as a
function of  velocity dispersion. The solid histogram shows the result for the
whole sample, while the  dotted and  dashed histograms show results for
radio-loud and radio-quiet clusters, respectively.} \label{fDet}
\end{figure}

As we showed in Section \ref{Ind_det},  only $\sim 25$ percent (159 out of 625)
of our  clusters are  individually detected in the RASS. The solid histogram in
figure \ref{fDet} shows the detected fraction as function of velocity
dispersion. As expected, the detected fraction is higher for  higher velocity
dispersion clusters. It increases from less than 15 per cent for groups with
$\sigma < 400\,\kms$ to more than 50 percent for clusters in the highest
velocity dispersion bin.

We find that 55 of the 134 radio-loud clusters and 88 of the 433 radio-quiet
clusters are detected. The detected fractions as function of velocity
dispersion for these two sub-samples are plotted in Fig. \ref{fDet}. As can be
seen, the detected fraction is systematically higher for radio-loud than for
radio-quiet clusters in all the velocity dispersion bins.

\subsubsection{$L_X-\sigma$ relation for clusters with individual X-ray detections}\label{Lsigma}

\begin{table*}
 \caption{Fitting parameters for various $L_X-\sigma$ relations. $a$
and $b$ are the fitting parameters of equation (\ref{fun}). $N$ is the
number of objects in each sample. For stacked samples, $N$ is given as
$N1/N2$, where $N1$ is the number of stacks while $N2$ is the total
number of clusters in the sample. $M$ is the median BCG stellar
mass. $b^\prime$ gives the zero-points obtained if the $L_X-\sigma$
relations are refit requiring the slope $a$ to be 2.97 for the low BCG
mass and low BCG mass radio-quiet samples and to be 4.40 for the other
stacked samples. }
\begin{tabular}{lccccc} \hline
Sample & $a$ & $b$ & $N$ & $M (\rm{M\odot})$ & $b^\prime$\\
\hline
Individual, all & $4.39\pm0.32$ & $-0.530\pm0.037$ & 159 & $1.82\times10^{11}$ & $-$ \\
Individual, radio-loud & $4.48\pm0.61$ & $-0.535\pm0.058$ & 55 & $2.09\times10^{11}$ & $-$ \\
Individual, radio-quiet & $4.50\pm0.44$ & $-0.530\pm0.056$ & 88 & $1.78\times10^{11}$ & $-$\\
Stack, radio-loud & $4.40\pm0.53$ & $-0.600\pm0.099$ & 8/134 & $1.66\times10^{11}$ & $-0.600\pm0.032$\\
Stack, control radio-quiet & $4.07\pm0.24$ & $-0.935\pm0.049$ & 8/134 & $1.35\times10^{11}$ & $-0.942\pm0.045$\\
Stack, low BCG mass & $2.88\pm0.17$ & $-1.150\pm0.055$ & 7/217 & $7.76\times10^{10}$ & $-1.151\pm0.033$\\
Stack, intermediate BCG mass & $4.58\pm0.56$ & $-0.712\pm0.116$ & 8/208  & $1.35\times10^{11}$ & $-0.703\pm0.037$\\
Stack, high BCG mass & $4.16\pm0.44$ & $-0.638\pm0.072$ & 8/200 & $2.14\times10^{11}$ & $-0.646\pm0.030$  \\
Stack, radio-loud, low BCG mass & $4.17\pm0.74$ & $-0.572\pm0.104$ & 6/67 & $1.20\times10^{11}$ & $-0.563\pm0.041$ \\
Stack, radio-loud, high BCG mass & $4.49\pm0.47$ & $-0.650\pm0.072$ & 6/67 & $2.14\times10^{11}$ & $-0.651\pm0.048$ \\
Stack, radio-quiet, low BCG mass & $3.07\pm0.21$ & $-1.160\pm0.053$ & 7/219 & $8.32\times10^{10}$ & $-1.162\pm0.032$ \\
Stack, radio-quiet, high BCG mass & $4.25\pm0.44$ & $-0.704\pm0.088$ &  8/214 & $1.74\times10^{11}$ & $-0.711\pm0.040$ \\
\hline
\end{tabular}\label{Tab_LS}
\end{table*}

\begin{figure}
  \includegraphics[width=84mm]{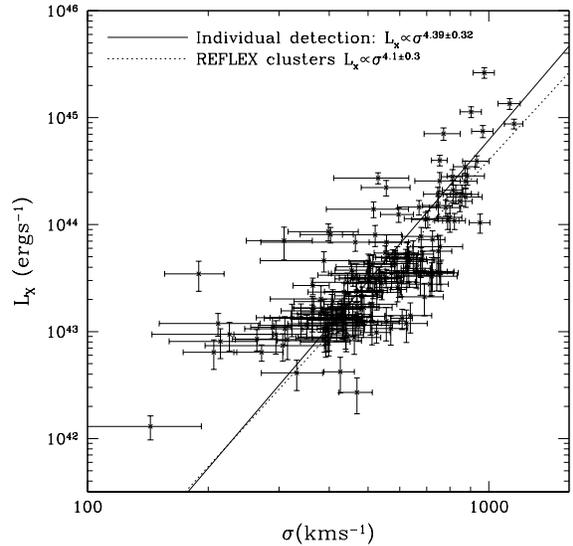}\\
\caption{The $L_X-\sigma$ relation of clusters that are individually detected
in the RASS. The solid line shows the linear fit of equation (\ref{LSAll}).
For comparison, the dotted line shows the fit obtained by Ortiz-Gil et
al.(2004) for REFLEX clusters. }
\label{LXBCG1}
\end{figure}

In figure~\ref{LXBCG1} we plot bolometric X-ray luminosity $L_X $ as function
of velocity dispersion $\sigma$ for clusters with individual X-ray detections.
$L_X$ is in units of $L_X/10^{44}\flux$($L_{44}$) and velocity dispersion in
units of $\sigma /500\kms$($\sigma_{500}$).

We use the BCES orthogonal distance regression method \citep{Akritas96} to fit
a linear relation between $\log L_{44}$ and $\log \sigma_{500}$,
\begin{equation}
\log L_{44} = a \log \sigma_{500} + b. \label{fun}
\end{equation}
This fitting method takes into account the observational errors on both
variables and the intrinsic scatter in the relation.  The determination of the
error on $L_X$ has been described in section \ref{Ind_det}, while the error on
$\sigma$ is given by \cite{Linden07}. The best fitting relation is
\begin{equation}
\log L_{44} = (4.39\pm0.32)\log \sigma_{500}+(-0.530\pm0.037), \label{LSAll}
\end{equation}
and is shown as solid line in Fig. \ref{LXBCG1}. The fitting parameters are
also listed in Table \ref{Tab_LS}. For comparison, we also plot the
$L_X-\sigma$ relation derived for REFLEX clusters by \cite{Ortiz-Gil04} using
the orthogonal distance regression method (dotted line). As we can see, the
slopes of two fitting relations are consistent with each other within the
1-$\sigma$ uncertainty. The zero-point of our relation is slightly higher. This
difference is caused by the fact that Ortiz-Gil et al. adopted a lower value of
$R_c$ for the extension correction [$R_c \propto L_x^{0.28}$, Equ. (6) of
\cite{Boehringer00}].

\begin{figure}
  \includegraphics[width=84mm]{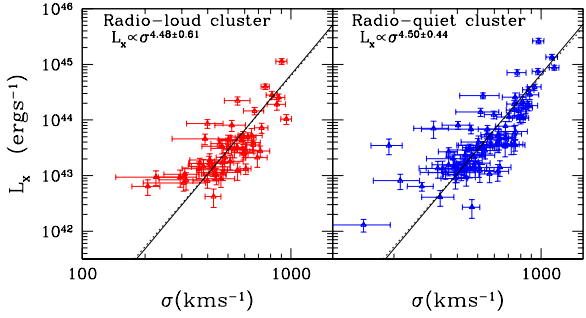}\vspace{-35mm}
\caption{The $L_X-\sigma$ relation of radio-loud (left panel) and radio-quiet
(right panel) subsamples of clusters.  The solid line in each panel is the best
linear fit [equations (\ref{LSRL}) and (\ref{LSRQ})], while the dotted line
shows the fit to the sample as a whole for comparison [equation (\ref{LSAll})].
} \label{LXBCG2}
\end{figure}

We show $L_X-\sigma$ relations for the sub-samples of {\it detected}
radio-loud and radio-quiet clusters in the left and right panels of figure
\ref{LXBCG2} respectively. The BCSE orthogonal regression method is again used
to fit a linear relation between $\log L_{44}$ and $\log \sigma_{500}$. The
results are
\begin{equation}
\log L_{44} = (4.48\pm0.61)\log \sigma_{500} + (-0.535\pm0.058), \label{LSRL}
\end{equation}
for radio-loud clusters and
\begin{equation}
\log L_{44} = (4.50\pm0.44)\log \sigma_{500} + (-0.530\pm0.056), \label{LSRQ}
\end{equation}
for radio-quiet clusters. These two relations are plotted as solid lines in
figure \ref{LXBCG2}. The relation for the cluster sample as a whole [Equ.
(\ref{LSAll})] is shown as a dotted line in each panel for comparison. Again,
the fitting parameters are listed in Table \ref{Tab_LS}.

Even though radio-loud clusters are more frequently detected in the X-ray than
radio-quiet objects, the two relations in Figure 4 are similar. As we
demonstrate in the next section, this is simply a selection effect. The
majority of individually detected clusters are just above the X-ray detection
limit in both samples. Since the redshift distributions of the two types of
cluster are similar (Fig. 1), this forces the mean $L_X-\sigma$ relations of
{\it detected} objects to be nearly the same.

\subsection{Results for the stacks}\label{Sta}
We stack the radio-loud and the radio-quiet clusters independently.
Clusters with unclear radio properties are excluded from this analysis.

Radio-loud clusters with $\sigma$ in the range $300-900\kms$  are stacked in
velocity dispersion bins with width  $100\kms$.  Smaller groups with $\sigma <
300\kms$ and bigger clusters with $\sigma > 900\kms$ are split into two
separate velocity dispersion bins. Our sample of 134 radio-loud clusters then
splits into 8 stacks.

As described above, we create a control sample of 134 radio-quiet clusters
selected to have the same $\sigma$ and $z$ distributions as the radio-loud
sample. This control sample is generated from the full sample of 433
radio-quiet clusters by picking the radio-quiet cluster that is most similar in
$z$ and $\sigma$ to each of the radio-loud clusters. These radio-quiet clusters
are then stacked in exactly the same way as the radio-loud clusters. More
specifically , for $i$-th radio-loud cluster that is included in the $k$-th
radio-loud stack, the corresponding $i$-th control radio-quiet cluster is
stacked into $k$-th radio-quiet stack. X-ray detections are obtained for all 16
stacks.

\subsubsection{Surface brightness profiles of the clusters}\label{SBprof}

\begin{figure*}
  \includegraphics[width=168mm]{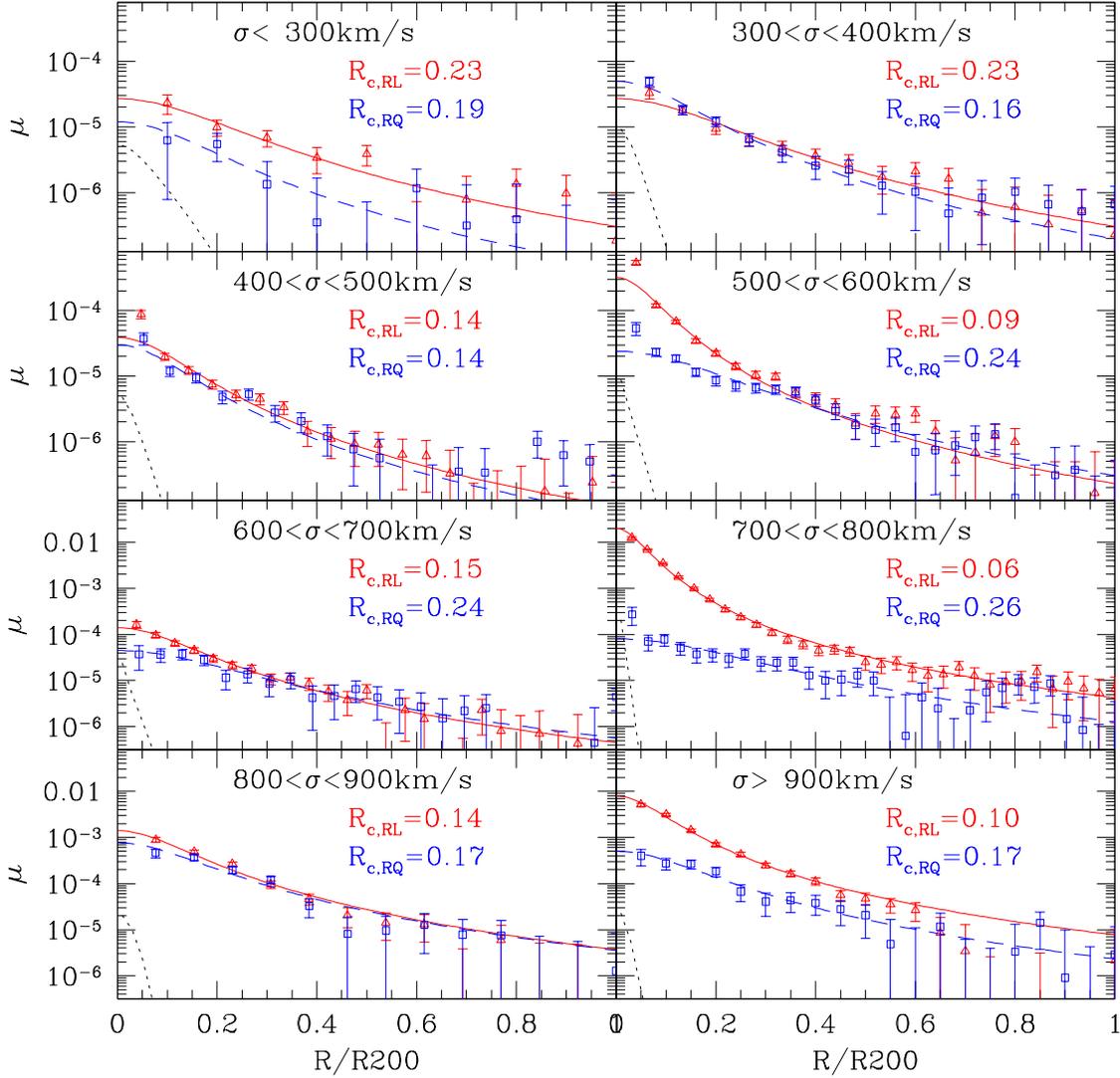}\\
\caption{Surface brightness profiles for stacks of clusters in 8 velocity
dispersion bins. Triangles show results for radio-loud clusters while squares
are for the control sample of radio-quiet clusters. The solid and dashed lines
show best fit $\beta$ models for radio-loud and radio-quiet clusters
respectively. The dotted curves show the contribution to the X-ray emission
predicted to come from the radio AGN itself (see the text in Section
\ref{AGNcont} for details). } \label{prof_stack}
\end{figure*}

We show radial surface brightness profiles for each of our eight different
velocity dispersion stacks in figure \ref{prof_stack}. The radio-loud clusters
are plotted as triangles while the control radio-quiet clusters are plotted as
squares. Surface brightness is given in units of photon counts per unit area
per second and is plotted as a function of the scaled radius $R/R_{200}$. The
error in the surface brightness in each radial bin is estimated from the
Poisson fluctuations of the photon counts. As we can see, after stacking, the
S/N of the surface brightness profiles in all the velocity dispersion bins is
sufficiently good to enable model fitting to be carried out.

We use a $\beta$ model [Equ. (\ref{betaprof})] to fit the surface brightness
profile for each stack. To reduce the number of free parameters, we fix
$\beta=2/3$ and estimate $R_c$ and $I_0$ by minimizing $\chi^2$. We show the
best fitting results as solid and dashed lines in each panel for radio-loud
and radio-quiet clusters respectively. The best fitting value of $R_c$ for
each profile is quoted as a label in each panel of Fig. \ref{prof_stack}.
Except for the two lowest velocity dispersion bins, radio-loud clusters have
more concentrated luminosity profiles (smaller $R_c$) than radio-quiet
clusters.

\begin{figure}
  \includegraphics[width=84mm]{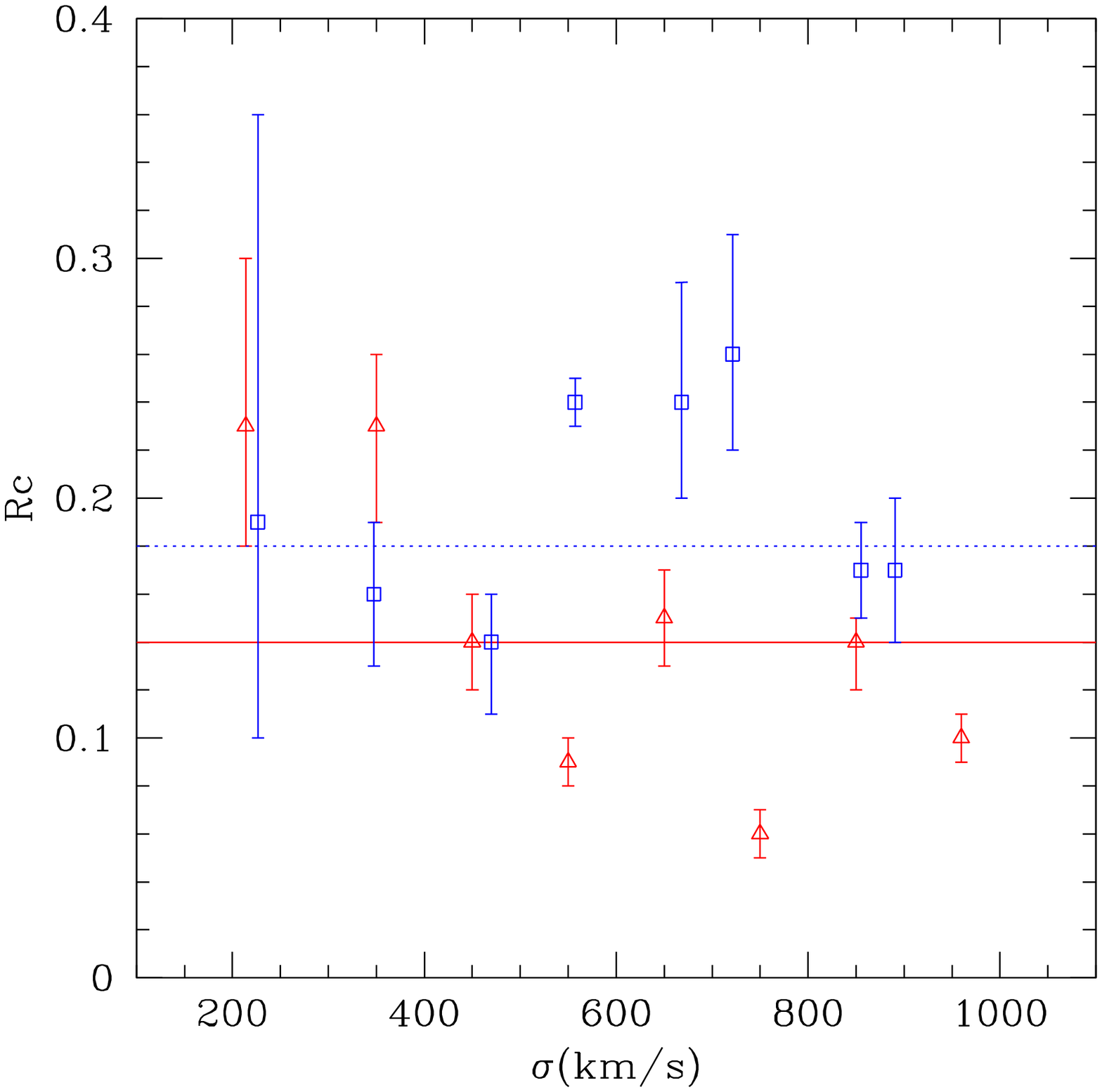}\\
\caption{The best fit value of $R_c$ for surface brightness profiles of
clusters stacked in 8 bins of velocity dispersion. Triangles and squares
represent radio-loud and radio-quiet clusters respectively.
$R_c=0.18R_{200}$ (solid line) and $R_c=0.14R_{200}$ (dotted line) are the
median values of $R_c$ for the radio-loud and radio-quiet clusters
respectively. } \label{Rc_sigma}
\end{figure}

We show the best fit values of $R_c$ as a function of velocity dispersion in
figure \ref{Rc_sigma}.  Radio-loud and radio-quiet clusters are represented by
triangles and squares, respectively. The median $R_c$ of the radio-loud
clusters is 0.14$R_{200}$, whereas the median for radio-quiet clusters is
0.18$R_{200}$. These two median values are shown as horizontal lines in Fig.
\ref{Rc_sigma}. The core radii of our stacked clusters are consistent with the
studies of individual clusters by \cite{Neumann99}.  These authors found
$R_c\sim0.1-0.2R_{200}$ for clusters with $\beta\sim2/3$. We note that
uncertainties in the centroids of individual cluster will broaden the X-ray
core of the stacked profile \citep[e.g.][]{Dai07}. This effect is not
significant in our study, however, since the centroids of most of our
individual detected clusters (which contribute the bulk of the flux of the
stacks) were identified before stacking(see Section \ref{Xcent}). We find that
if we stack {\it only} the clusters with individual detections, the median
$R_c$ values are $0.14R_{200}$ and $0.17R_{200}$ for radio-loud and radio-quiet
clusters respectively. The point-spread function of the ROSAT telescope also
broadens the estimated core radii of the clusters, of course, particularly for
low velocity dispersion clusters which typically have small angular size.

\subsubsection{$L_X-\sigma$ relation for the  stacks}\label{LSstack}

\begin{figure}
\includegraphics[width=84mm]{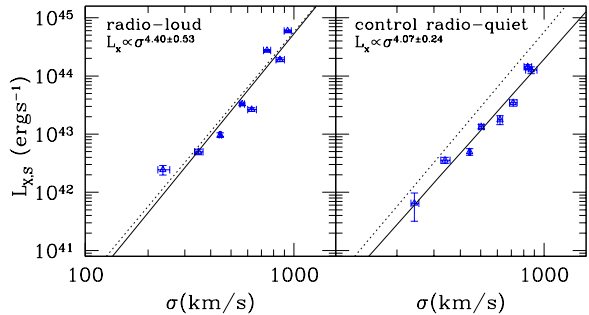}\vspace{-35mm}
\caption{The average X-ray luminosity $L_{X,S}$ as a function of average
velocity dispersion $\sigma_{S}$  for stacked clusters. The left and  right
panels show  results for  radio-loud and radio-quiet clusters respectively. The
solid lines show the best linear fits to the relation between $\log L_{X,S}$ and
$\log\sigma_{S}$ [Equ. (\ref{SLSRL}) and Equ. (\ref{SLSRQ})]. The $L_X-\sigma$
relation for all clusters with individual X-ray detections [Equ. (\ref{LSAll})]
is shown as a dotted line in each panel for comparison. }\label{LsigmaS}
\end{figure}

In figure \ref{LsigmaS}, we show the weighted average X-ray bolometric
luminosity $L_{X,S}$[Equ. (\ref{Lstack})] as a function of the average
velocity dispersion $\sigma_{S}$ for our stacks of radio-loud and radio-quiet
clusters.  We use the BCSE orthogonal regression method to fit linear
relations between $\log L_{X,S}$ and $\log \sigma_{S}$ weighted by the errors
on both $L_{X,S}$ and $\sigma_{S}$.  The error on $\sigma_{S}$ is estimated
from the error on the mean value of $\sigma$ of the stacked clusters. Our
result is
\begin{equation}
\log L_{44} = (4.40\pm0.53)\log \sigma_{500} + (-0.600\pm0.099) \label{SLSRL}
\end{equation}
for radio-loud clusters and
\begin{equation}
\log L_{44} = (4.07\pm0.24)\log \sigma_{500} + (-0.935\pm0.049) \label{SLSRQ}
\end{equation}
for radio-quiet clusters. These relations are shown as solid lines in the left
and right panels of Fig. \ref{LsigmaS} for radio-loud and radio-quiet clusters
respectively. The parameters of the fits are also listed in Table
\ref{Tab_LS}. The $L_X-\sigma$ relation for the individually detected clusters
[Equ. (\ref{LSAll})] is shown as a dotted line in each panel for comparison.
The $L_X-\sigma$ relation of the stacks of radio-loud clusters is very close
to that for all individual detections (dotted line). This is a coincidence.  If
we stack only the individually detected clusters, we find a significantly
higher $L_X$ at given $\sigma$ than predicted by Equ. (\ref{LSAll}) (the
dotted line). This is because, at given $\sigma$, the mean $L_X$ of the
stacked objects is higher than their median $L_X$. The latter is what is
approximated by a linear fit in $\log \sigma$ -- $\log L_X$ space.

The slopes of the $L_X-\sigma$ relations for the radio-loud and (control)
radio-quiet clusters are consistent within the 1-$\sigma$ error, but their
zero-points differ significantly. At given velocity dispersion, the average
X-ray luminosity of radio-loud clusters is systematically higher than that of
radio-quiet clusters. To demonstrate the significance of this effect, we fix
the slope of both relations to be 4.17 [the error-weighted mean of the slopes
in Equ. (\ref{SLSRL}) and (\ref{SLSRQ})], and then re-estimate their
zero-points.  The results are ($-0.591\pm0.031$) for radio-loud clusters and
($-0.937\pm0.044$) for radio-quiet clusters. At given velocity dispersion, the
X-ray luminosity of radio-loud clusters is on average 2.2 times higher than
that of radio-quiet clusters. The difference is significant at 6.4$\sigma$ and
is consistent with our earlier result that the fraction of radio-loud clusters
with individual X-ray detections is substantially higher than the
corresponding fraction for radio-quiet clusters (Section \ref{frac}). The
difference is also comparable to the difference in normalization between
cooling core and non-cooling core clusters found by \cite{Chen07} using the
$L_X-M$ relation ($M$ is cluster mass) .

\section{Discussion}
The results presented above demonstrate that the X-ray properties of the ICM
are correlated with the radio properties of the central BCGs. Clusters with
radio-loud BCGs are more frequently detected in X-ray images. When we stack
radio-loud and radio-quiet clusters with similar velocity dispersions and
redshifts, we find that radio-loud clusters have more concentrated surface
brightness profiles and higher average X-ray luminosities than their
radio-quiet counterparts.

Up to now, we have not considered X-ray emission from the central radio AGN
itself. If this emission were comparable to the X-ray emission from the ICM,
all our results might be explained without any need to invoke a correlation
between the radio AGN and the state of the intracluster gas. We estimate the
X-ray emission from the AGN itself in Section \ref{AGNcont}.

Furthermore, we have not yet considered how X-ray and radio properties
correlate with the stellar properties of BCGs.  As shown by Best et al.(2005,
2007), radio-loud AGN occur more frequently in higher stellar mass BCGs, which
also tend to be found in more X-ray luminous clusters\citep{Edge91}. The
correlation between BCG stellar mass and cluster X-ray luminosity is
two-fold. On the one hand, more massive clusters typically host higher mass
BCGs(see top panel of Fig. \ref{BCG_RLRQ}).  On the other hand, at given
cluster mass (velocity dispersion), clusters with higher mass BGCs tend to be
more regular and more concentrated\citep{Bautz70}, and such clusters tend to
have higher X-ray luminosities\citep{David99,Ledlow03}. This morphology -- BCG
mass dependence appears related to the presence of the cool
cores\citep{Edge91}.  Thus, we may expect the mean stellar masses of the BCGs
to differ between the radio-loud and radio-quiet clusters we have stacked
above, and we may expect clusters stacked as a function of BCG mass to show
similar differences in X-ray properties to our radio-loud and radio-quiet
samples. In Section \ref{BCGmass}, we compare the stellar properties of the
BCGs in our radio-loud and radio-quiet samples, and we stack clusters also as
a function of BCG properties.

\subsection{The X-ray emission from radio AGN}\label{AGNcont}
\begin{figure}
\includegraphics[width=84mm]{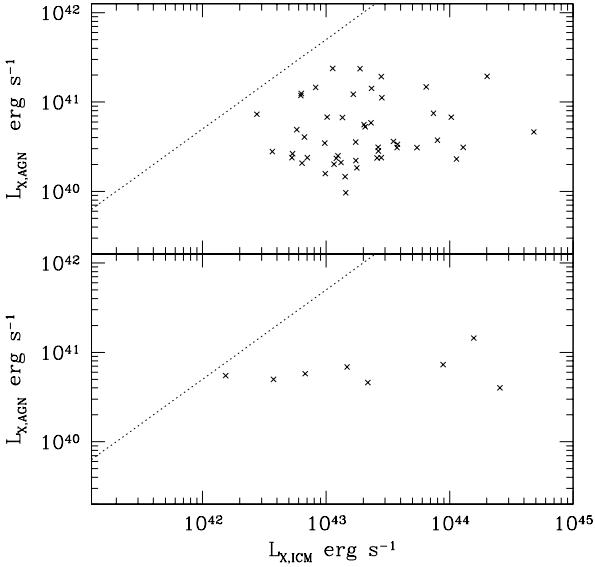}\\
\caption{ Comparison of the predicted X-ray emission from radio
AGN($L_{X,AGN}$) to the total observed cluster emission ($L_{X,ICM}$) for
our sample of radio-loud clusters. The AGN luminosity is estimated from the
$L_X-L_R$ relation of Brinkmann et al.(2000). The top panel gives results for
individually detected clusters, the bottom panel for our eight stacks of
clusters. The dotted lines represent the relation
$L_{X,AGN}=0.05L_{X,ICM}$. }\label{LxAGN}
\end{figure}

The X-ray luminosities of radio-loud AGN $L_{X,AGN}$ are correlated with their
radio luminosities $L_R$ \citep[e.g.][]{Brinkmann00, Merloni03}. Using the
$L_{X,AGN}-L_R$ correlation for radio-loud AGN from the study of Brinkmann et
al. (2000), we estimate X-ray luminosities for our sample of radio AGN
($L_{X,AGN}$) in the ROSAT $0.1-2.4\kev$ band. Here we assume that the
distributions of radio flux and of X-ray photon energy are power laws, $S_\mu
\sim \mu^{-0.5}$ and $N(E)\sim E^{-2}$ \citep[see][]{Brinkmann00}. We then
compare our estimated values of $L_{X,AGN}$ with the $0.1-2.4\kev$ X-ray
luminosities $L_{X,ICM}$ measured for our clusters within the radius $R_X$. The
results are shown in figure \ref{LxAGN}. The top panel shows results for
individually detected clusters; the bottom panel shows corresponding results
for the stacked clusters.  For the stacks, we estimate $L_{X,AGN}$ as the
weighted average of the estimated X-ray luminosities of the central radio AGN.

As can be seen, the estimated X-ray luminosities of the radio AGN are
typically a small fraction ($<5$ percent) of the total measured X-ray
emission. The enhancement we measure in the X-ray luminosity of radio-loud
clusters is more than a factor of 2 (Section \ref{Sta}). Thus, contamination
of the X-ray emission by the radio AGN is at most a small part of the effect
we measure.

One might also ask whether X-ray emission from the radio AGN might explain the
more concentrated X-ray surface brightness profiles seen in figure
\ref{prof_stack}. The dotted curve in each panel shows the predicted
contribution of the radio AGN to the X-ray surface brightness profile. The
$L_X-L_R$ relation of \cite{Brinkmann00} was again used to estimate the X-ray
luminosity of each radio AGN and a Gaussian PSF with FWHM = 1 arcmin was used
to predict its count rate profile. These count rate profiles were then scaled
and stacked in order to calculate the average AGN contribution to the profile
of the stack.  As we can see, this contribution is negligible even in the
lowest velocity dispersion bin.

\subsection{The stellar properties of BCGs}\label{BCGmass}

In this section we investigate the relation between our results and
the stellar properties of our BCGs, as characterised by their stellar
mass and their concentration.  The concentration is defined as
$c\equiv R_{90}/R_{50}$. where $R_{90}$ and $R_{50}$ are the radii
including 90 and 50 percent of the flux from a galaxy. It can be used
as an indicator of galaxy morphology\citep{Shimasaku01}. For
early-type galaxies, the stellar mass correlates closely with the mass
of the central supermassive black hole.

\begin{figure}
\includegraphics[width=84mm]{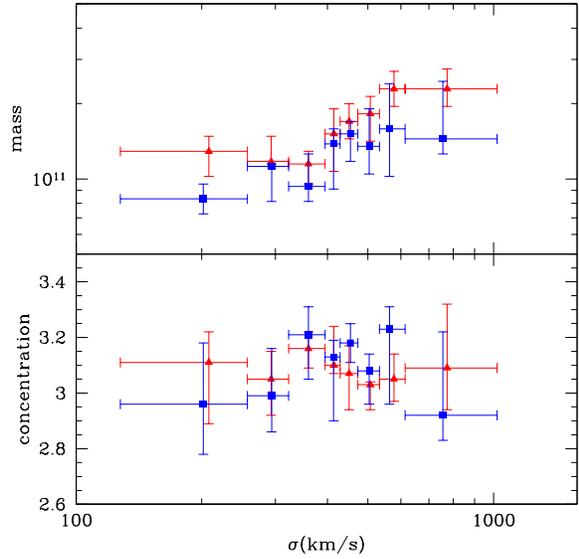}\\
\caption{Median stellar mass (upper panel) and concentration (lower
panel) for BCGs as function of the velocity dispersion of their host
cluster.  The red triangles and blue squares represent radio-loud and
control radio-quiet BCGs, respectively. Horizontal error-bars show the
range of the velocity dispersion bins, while vertical error-bars
indicate the 32 to 68 percent ranges of the
distributions. }\label{BCG_RLRQ}
\end{figure}

We first divide our radio-loud clusters into 8 velocity dispersion bins
containing equal numbers of objects. The median stellar mass and concentration
are then calculated for both radio-loud and control radio-quiet clusters in
these 8 bins. In figure \ref{BCG_RLRQ}, we show these median values as a
function of velocity dispersion.  For both properties red triangles represent
radio-loud clusters, while blue squares represent the radio-quiet clusters. The
horizontal error-bars show the range of velocity dispersion for each bin, while
the vertical error-bars link the 32 and 68 percentiles of the distribution in
concentration and stellar mass.

Clearly, the stellar masses of BCGs correlate with the velocity
dispersions of their host clusters; higher velocity dispersion
clusters tend to have higher stellar mass BCGs. In addition, at given
velocity dispersion, the stellar mass of radio-loud BCGs is
systematically higher than that of radio-quiet objects. This mirrors
the earlier finding of Best et al.(2005, 2007) that higher stellar
mass galaxies, both BCGs and non-BCGs, are more likely to host
radio-loud AGN, but that at fixed stellar mass the BCGs are more
likely to be radio-loud than the non-BCGs.  Concentration values show
no consistent trends with velocity dispersion or radio activity and
are typically $c\sim3$, in the range expected for a galaxy with a
de Vaucouleurs profile.

Given that the radio-loud AGN are biased towards higher stellar mass
galaxies, it is natural to ask whether the higher $L_X$ of our
radio-loud clusters simply reflects the larger stellar masses of their
BCGs.  To investigate this issue, we first study the dependence of the
$L_X-\sigma$ relation on BCG stellar mass, independent of BCG
radio activity.

As before, we first divide our 625 clusters of galaxies into 8 velocity
dispersion bins. We take the range of each bin to be the same as in Fig.
\ref{LsigmaS}. We then divide the clusters in each bin into three equal
sub-samples according to the stellar mass of their BCGs. Combining velocity
dispersion bins, gives us three sub-samples, which we refer to as the low,
intermediate and high mass BCG samples. The median stellar mass of BCGs in
these three samples are $7.76\times10^{10}M_\odot$, $1.35\times10^{11}M_\odot$
and $2.14\times10^{11}M_\odot$. For comparison, the median stellar mass of the
BCGs in our radio-loud and control radio-quiet cluster samples are
$1.66\times10^{11}M_\odot$ and $1.35\times10^{11}M_\odot$, respectively. Thus,
the high BCG mass sample has even larger median BCG mass than our radio-loud
sample, while the low BCG mass sample has even lower median BCG mass than our
control radio-quiet sample. The intermediate BCG mass sample have nearly the
same median BCG mass as the control radio-quiet sample.

\begin{figure}
\includegraphics[width=84mm]{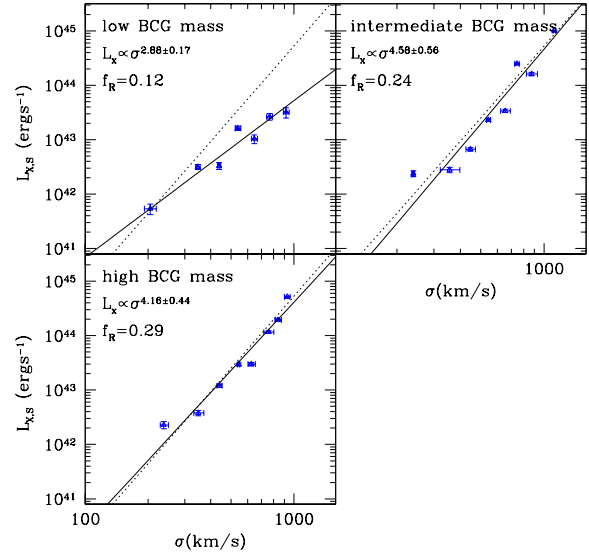}\\
\caption{The $L_X-\sigma$ relation for stacks of clusters with high,
intermediate and low BCG stellar masses (top left, top right and
bottom left panels respectively). The solid lines in each panel show
the best linear fits to the relation between $\log L_X$ and $\log
\sigma$. The $L_X-\sigma$ relation for the stacks of radio-loud
clusters [Equ. (\ref{SLSRL})] is shown as dotted line in each panel
for comparison.}\label{LSmass}
\end{figure}

For each BCG mass sub-sample, we then combine all the clusters in each velocity
dispersion bin into a single stack. For the low BCG mass sample, the X-ray
detection of the velocity dispersion bin $\sigma>900\kms$ is not significant,
so we combine all the clusters with $\sigma>800\kms$ into one stack, thus
ending up with seven stacks for the low BCG mass sample and eight stacks for
the intermediate and high BCG mass samples.  We show the resulting $L_X-\sigma$
relations in the three panels of Fig. \ref{LSmass}.  As before, we use the BCSE
orthogonal regression method to fit linear relations between $\log L_X$ and
$\log \sigma$ weighted by the error on both $L_X$ and $\sigma$.  The fitting
relations are shown as solid lines each panel of Fig. \ref{LSmass}, and the
corresponding fitting parameters are listed in Table 1. The $L_X-\sigma$
relation for our stacks of radio-loud clusters [Equ.  (\ref{SLSRL})] is shown
as a dotted line in each panel for reference.

The X-ray luminosity of the clusters shows a very interesting
dependence on the stellar mass of the BCGs. The high and intermediate
BCG mass samples have have $L_X-\sigma$ relations which are both
essentially identical to that of the radio-loud clusters. The low BCG
mass sample, on the other hand, has an $L_X-\sigma$ relation which is
significantly different. Clusters in the higher velocity dispersion
bins are substantially less luminous than those in the other two
samples. This results in a much shallower slope, $2.88\pm0.17$, than
the value $\sim 4$ found in all the other cases.

The fractions of radio-loud clusters are, as expected, higher in the
samples with higher mass BCGs. The numbers for the low, intermediate
and high mass samples are 12, 24 and 29 percent, respectively. We note
that radio-loud fraction in the intermediate and high BCG mass samples
do not differ very much, while the low BCG mass sample has
significantly fewer radio-loud objects. Together, all these results
imply that the stellar mass of BCGs is strongly coupled both with
their radio activity and with the properties of the ICM. We now
explore this connection further by examining the BCG stellar mass
dependence of the $L_X-\sigma$ relation separately for radio-loud and
radio-quiet clusters.

We split our radio-loud sample into two sub-samples with similar
velocity dispersion and radio properties, but with systemically
different BCG stellar masses. In detail, we separate the 134
radio-loud clusters into 67 pairs with very similar velocity
dispersion and BCG radio luminosity. We then take the clusters with
the higher BCG stellar mass in each pair to build the high BCG mass
sub-sample, and the clusters with lower BCG stellar mass to build the
low BCG mass sample. We split the 67 high BCG mass clusters into 6
velocity dispersion bins and make 6 stacks. Clusters with $\sigma$ in
the range $300-600\kms$ are split into 3 equal bins with width
$100\kms$; clusters with $600<\sigma<800\kms$ make up another
stack. The remaining groups with $\sigma<300\kms$ and clusters with
$\sigma>800\kms$ make up the two remaining stacks. For the low BCG
mass sample, we make 6 corresponding stacks. For each pair of
radio-loud clusters, if the high BCG mass object is binned into $j$th
stack, the low BCG mass object is binned into $j$th stack also.

We also split the radio-quiet clusters into two sub-samples with
differing BCG stellar mass. As for the whole sample, we first bin the
clusters into 8 velocity dispersion bins. In each velocity dispersion
bin, we then separate the clusters into equal low BCG mass and high
BCG mass sub-samples. For each sub-sample, we combine all the clusters
in each velocity dispersion bin into a single stack. For the low BCG
mass sample, we combine the clusters in the two highest velocity
dispersion bins ($800\kms<\sigma<900\kms$ and $\sigma>900\kms$), since
the significance of the X-ray detection of the $\sigma>900\kms$ bin is
too low.

\begin{figure}
\includegraphics[width=84mm]{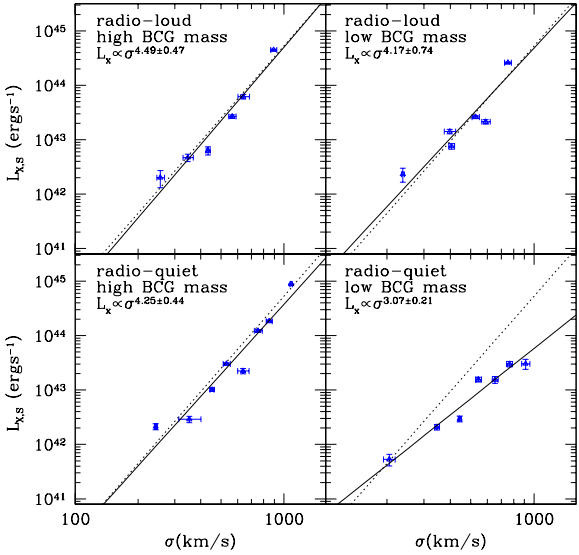}\\
\caption{The $L_X-\sigma$ relation for stacks of low BCG mass and high
BCG mass sub-samples of radio-loud and radio-quiet clusters. The
results for radio-loud and radio-quiet clusters are shown on the top
and bottom rows, respectively. High BCG mass sub-samples are shown on
the left and low BCG mass sub-samples on right. The solid lines in
each panel show the best linear fits to the relations between $\log
L_X$ and $\log \sigma$. The $L_X-\sigma$ relation for stacks of the
full sample of radio-loud clusters [Equ. (\ref{SLSRL})] is shown as
dotted line in each panel for reference.}\label{RLRQ_mass}
\end{figure}

We obtain significant X-ray detections for all stacks in these four
samples. The resulting $L_X-\sigma$ relations are shown in figure
\ref{RLRQ_mass}. The two sub-samples of radio-loud clusters are shown
in the top two panels, with the high BCG mass sample on the left and
the low BCG mass sample on the right. The two radio-quiet sub-samples
are shown in the same order in the bottom two panels.  We use the BCSE
orthogonal regression method to fit linear relations to the $\log L_X$
and $\log \sigma$ data in all these panels, weighting by the errors in
both $L_X$ and $\sigma$. The fitting relations are shown as solid
lines in each panel and the corresponding parameters are listed in
Table 1. The $L_X-\sigma$ relation for the radio-loud stacks of the
full sample [Equ.  (\ref{SLSRL})] is again shown as a dotted line in
each panel for reference.

Three of these four subsamples show very similar $L_X-\sigma$
relations, all of which are consistent with the relation for the full
radio-loud sample. Only the radio-quiet subsample with low BCG mass
shows a significantly different relation which in turn is very similar
to that shown in Fig. \ref{LSmass} for the subsample with lowest BCG
mass, independent of radio activity. Thus it seems that the
$L_X-\sigma$ is similar for all clusters except those for which are
{\it both} radio-quiet {\it and} have an unusually low mass BCG. Among
radio-loud clusters we detect no dependence of the relation on BCG
stellar mass. The difference between radio-loud and radio-quiet
clusters appears to be due entirely to the presence of a subsample of
relatively high velocity dispersion radio-quiet clusters which have
both low BCG stellar mass and low X-ray luminosity. These might
plausibly be systems which have not yet fully collapsed, so that both
their X-ray luminosity and their BCG mass are typical of those usually
found in lower mass relaxed systems.

To look more carefully for off-sets between our various subsamples we
have refit the data plotted in Figs \ref{LSmass} and \ref{RLRQ_mass}
fixing the slopes $a$ to be 2.97 for the low BCG mass and low BCG mass
radio-quiet panels and 4.40 for the other five panels.  The resulting
zero-points $b^\prime$ are listed with their formal uncertainties in
Table 1. The relations in each group are formally consistent within
their errors, so we see no clear evidence justifying a more complex
interpretation of the data. The most significant and suggestive
difference is that the full radio-loud cluster sample is about 30\%
more X-ray luminous than the sample of high BCG mass radio-quiet
clusters, even though the latter sample has a slightly larger median
BCG mass. This result is significant at just over the $2\sigma$ level,
suggesting that even at fixed cluster velocity dispersion and BCG
mass, radio sources prefer to live in more X-ray luminous clusters.

\section{Conclusions}

We have used the ROSAT All Sky Survey to study the X-ray properties of
a sample of 625 groups and clusters of galaxies selected from the
Sloan Digital Sky Survey. We focus on the $L_X-\sigma$ relation of
these groups and clusters, and we study whether this relation
depends on the radio properties of the central galaxy (BCG). A cluster
is termed `radio-loud' if its central BCG is a radio-loud AGN, and
`radio-quiet' if the BCG is not detected at radio wavelengths. We find
that the fraction of clusters with individual X-ray detections depends
strongly on whether the BCG is radio-loud. Radio-loud clusters are
detected more frequently than radio-quiet clusters of the same
velocity dispersion and redshift.

The $L_X-\sigma$ relations for individually detected radio-loud and
radio-quiet clusters are very similar, but this is purely a selection bias.
The majority of detected clusters are just above the X-ray detection limit in
both samples. Since the redshift distribution at each velocity dispersion is
similar for the two types of cluster, the mean relations for {\it detected}
objects are forced to be similar.

By stacking the X-ray images of clusters with similar velocity dispersion, we
studied the {\em average} X-ray luminosities and surface brightness profiles of
our clusters as function of velocity dispersion. The average X-ray luminosities
of radio-loud clusters are systematically higher and their luminosity profiles
are more concentrated than those of radio-quiet systems. X-ray emission from
the radio AGN itself is by far insufficient to explain this doubling of the
X-ray luminosity.  Our results demonstrate convincingly and quantitatively that
the X-ray properties of the intracluster gas correlate with the presence of a
central radio source in the way first suggested by Burns (1990).

The stellar masses of the BCGs also correlate with their radio properties;
radio-loud clusters tend to mave more massive BCGs than radio-quiet
clusters of the same velocity dispersion. Those clusters of given
velocity dispersion which have unusually low mass BCGs tend to be both
underluminous in X-rays and radio-quiet. This effect is particularly
pronounced for large velocity dispersion clusters.  Among radio-loud
clusters we find no dependence of X-ray luminosity on BCG stellar
mass. These results can be summarised by saying that the only clear
dependence of X-ray properties on BCG properties is that high velocity
dispersion clusters in which the BCG is both low mass and radio-quiet
tend to be several times less X-ray luminous than otherwise similar
clusters in which the BCG is massive and/or radio-loud.

Clearly a high central hot gas density is needed for effective
fuelling of the radio source. If the radio activity is in turn able to
heat the surrounding gas and cause its re-expansion, the feedback
cycle needed to control the growth of the central galaxy could
be established. The conditions which lead to effective AGN fuelling
are clearly related to the presence of a massive central galaxy,
although the exact causal relationship between BCG growth, AGN
activity and ICM structure remains to be clarified.

\section*{Acknowledgments}
We thank the anonymous referee for posing questions which
significantly clarified the analysis in this paper.  SS acknowledges
the financial support of MPG for a visit to MPA. This project is
supported by the Knowledge Innovation Program of the Chinese Academy
of Sciences, NSFC10403008, NKBRSF2007CB815402, and Shanghai Municipal
Science and Technology Commission No. 04dz\_05905.

\bibliographystyle{mn2e}

\clearpage \onecolumn

\begin{appendix}
\section{The clusters of galaxies individually detected in RASS}
Here we list the basic properties of 142 clusters of galaxies individually
detected in RASS. The description of the columns is as follows.

\begin{itemize}
 \item [-] column 1: the ID of cluster in SDSS C4 cluster  catalogue(http://www.ctio.noao.edu/~chrism/C4/)
 \item [-] column 2: the Right Ascension(J2000) of the BCG in decimal degrees
 \item [-] column 3: the Declination(J2000) of the BCG in decimal degrees
 \item [-] column 4: the Right Ascension(J2000) of the X-ray centre in decimal degrees
 \item [-] column 5: the Declination(J2000) of the X-ray centre in decimal degrees
 \item [-] column 6: the redshift of the BCG
 \item [-] column 7: the virial radius $R_{200}$ in unit of arcmin
 \item [-] column 8: the velocity dispersion in unit of $\kms$
 \item [-] column 9: the flux within $R_X$  in the energy band 0.5-2.0kev, in unit of $10^{-12}$ erg $\rm{s}^{-1}\rm{cm}^{-2}$
 \item [-] column 10: the X-ray extension radius $R_X$, in unit of arcmin
 \item [-] column 11: the fractional error on count rate(flux) within $R_X$
 \item [-] column 12: the extension correction factor $f_E$[equation (\ref{fe})]
 \item [-] column 13: the bolometric X-ray luminosity after the extension, in unit of $10^{44}$ erg $\rm{s}^{-1}$
 \item [-] column 14: the classification of radio properties of BCG, 1 for the
clusters with BCG to be radio-loud AGN, 0 for the clusters with BCG to be
radio-quiet, $-1$ for  others
\end{itemize}

\begin{longtable}{lcccccccccccccc}
\hline
ID & RA & Dec & $\rm{RA_X}$ & $\rm{Dec_X}$ & $z$ & $R_{200}$ & $\sigma$  & $f_X$ & $R_X$ & Err & $f_E$ & $L_{44}$ & R-class\\
(1) & (2) & (3) & (4) & (5) & (6) & (7) & (8) & (9) & (10) & (11) & (12) & (13) & (14) \\
\hline
\endhead
  1000  &  202.54301  &   -2.10501  &       $-$  &       $-$  &    0.0867  &    14.2  &   647.8  &     1.553  &     7.0  &     0.308  &     1.186  &     0.529  &    1 \\
  1048  &  205.54018  &    2.22721  &  205.5251  &    2.2276  &    0.0774  &    20.5  &   827.5  &     5.339  &     9.0  &     0.147  &     1.324  &     1.950  &    0 \\
  1066  &  202.79596  &   -1.72731  &  202.8011  &   -1.7162  &    0.0854  &    18.2  &   814.2  &     6.297  &    18.0  &     0.218  &     1.002  &     2.102  &    0 \\
  1001  &  208.27667  &    5.14974  &  208.3057  &    5.2152  &    0.0794  &    18.0  &   746.4  &     6.316  &    11.0  &     0.116  &     1.146  &     1.924  &    0 \\
  1002  &  159.77759  &    5.20977  &       $-$  &       $-$  &    0.0690  &    22.3  &   800.4  &     4.222  &    10.0  &     0.241  &     1.311  &     1.168  &    0 \\
  1004  &  184.42136  &    3.65581  &  184.4206  &    3.6600  &    0.0774  &    23.9  &   966.0  &    19.892  &    14.0  &     0.055  &     1.165  &     7.462  &    0 \\
  1017  &  182.57005  &    5.38603  &  182.5764  &    5.3857  &    0.0769  &    14.8  &   596.0  &     5.263  &    10.0  &     0.111  &     1.109  &     1.242  &    0 \\
  1069  &  184.71817  &    5.24567  &       $-$  &       $-$  &    0.0764  &    18.1  &   721.3  &     1.161  &     8.0  &     0.323  &     1.320  &     0.365  &    0 \\
  1087  &  183.73737  &    5.04247  &  183.7515  &    5.0156  &    0.0782  &    11.4  &   465.0  &     2.945  &     7.0  &     0.135  &     1.144  &     0.684  &    0 \\
  1212  &  148.42239  &    1.70070  &  148.4196  &    1.7052  &    0.0977  &     7.5  &   388.3  &     1.424  &     6.0  &     0.189  &     1.042  &     0.459  &    1 \\
  1006  &  191.30367  &    1.80480  &       $-$  &       $-$  &    0.0477  &    13.9  &   340.3  &     1.950  &    10.0  &     0.327  &     1.085  &     0.135  &    0 \\
  1044  &  194.67288  &   -1.76146  &  194.6711  &   -1.7589  &    0.0837  &    17.6  &   771.0  &    19.671  &    10.0  &     0.061  &     1.178  &     7.064  &    0 \\
  1058  &  195.71906  &   -2.51635  &  195.7177  &   -2.5096  &    0.0831  &    17.2  &   748.7  &     3.737  &     7.0  &     0.160  &     1.381  &     1.511  &    0 \\
  1016  &  175.29919  &    5.73480  &  175.2950  &    5.7151  &    0.0983  &    12.7  &   660.1  &     0.799  &     5.0  &     0.325  &     1.296  &     0.391  &    1 \\
  1041  &  179.37073  &    5.08906  &       $-$  &       $-$  &    0.0758  &    17.1  &   678.2  &     2.836  &     9.0  &     0.176  &     1.217  &     0.774  &    0 \\
  1011  &  198.05661  &   -0.97449  &  198.0540  &   -0.9832  &    0.0847  &    14.2  &   631.4  &     1.115  &     5.0  &     0.328  &     1.511  &     0.453  &    0 \\
  1028  &  199.13568  &    0.87025  &  199.1394  &    0.8830  &    0.0796  &     8.7  &   363.5  &     0.525  &     3.0  &     0.331  &     1.535  &     0.156  &    0 \\
  1047  &  197.32954  &   -1.62253  &  197.3326  &   -1.6232  &    0.0829  &    12.0  &   521.1  &     3.234  &     9.0  &     0.197  &     1.055  &     0.803  &    1 \\
  1189  &  201.57338  &    0.22150  &  201.5723  &    0.2214  &    0.0829  &    11.9  &   516.6  &     5.114  &     7.0  &     0.122  &     1.163  &     1.394  &    0 \\
  1372  &  201.76564  &    1.33573  &  201.7738  &    1.3475  &    0.0804  &    16.1  &   677.4  &     1.569  &     9.0  &     0.298  &     1.186  &     0.471  &    0 \\
  1013  &  227.10735  &   -0.26629  &  227.1060  &   -0.2585  &    0.0906  &    15.7  &   747.9  &     0.830  &     4.0  &     0.327  &     1.944  &     0.564  &    0 \\
  1151  &  226.68803  &   -1.23171  &  226.6998  &   -1.2309  &    0.0710  &    13.5  &   500.4  &     1.928  &    10.0  &     0.308  &     1.078  &     0.349  &    0 \\
  1355  &  227.88237  &    1.76388  &  227.8839  &    1.7574  &    0.0390  &    19.6  &   391.7  &     2.039  &    11.0  &     0.261  &     1.184  &     0.105  &    0 \\
  1014  &  220.17848  &    3.46542  &  220.1600  &    3.4738  &    0.0269  &    33.4  &   459.0  &    10.640  &    14.0  &     0.092  &     1.359  &     0.317  &    0 \\
  1025  &  153.40948  &   -0.92541  &  153.4424  &   -0.8592  &    0.0451  &    34.0  &   789.8  &     8.720  &    13.0  &     0.092  &     1.434  &     1.085  &    0 \\
  1075  &  153.43707  &   -0.12022  &  153.4424  &   -0.1068  &    0.0944  &    17.6  &   875.5  &     3.496  &    10.0  &     0.163  &     1.178  &     1.814  &    0 \\
  1167  &  154.45268  &   -0.03077  &  154.4389  &   -0.0330  &    0.0638  &    12.2  &   405.9  &     0.732  &     5.0  &     0.296  &     1.379  &     0.130  &    0 \\
  1020  &  214.39804  &    2.05322  &       $-$  &       $-$  &    0.0540  &    21.7  &   605.2  &     2.546  &    13.0  &     0.248  &     1.115  &     0.292  &    1 \\
  1200  &  216.19754  &    2.66442  &  216.1944  &    2.6688  &    0.0543  &    21.1  &   592.4  &     1.911  &     8.0  &     0.232  &     1.318  &     0.259  &    1 \\
  1032  &  218.49640  &    3.77800  &       $-$  &       $-$  &    0.0290  &    38.4  &   569.8  &     2.471  &    13.0  &     0.260  &     1.552  &     0.107  &    0 \\
  1039  &  186.87810  &    8.82456  &  186.8716  &    8.8272  &    0.0897  &    17.9  &   846.1  &     3.487  &     9.0  &     0.160  &     1.243  &     1.655  &    0 \\
  1042  &  228.80879  &    4.38621  &  228.8193  &    4.3951  &    0.0980  &    16.6  &   856.8  &     3.443  &     8.0  &     0.183  &     1.194  &     1.914  &    1 \\
  1351  &  228.79631  &    3.84851  &  228.7980  &    3.8472  &    0.0786  &    11.4  &   468.1  &     1.327  &     5.0  &     0.243  &     1.237  &     0.337  &    1 \\
  1043  &  168.33385  &    2.54667  &  168.3346  &    2.5354  &    0.0743  &    10.4  &   403.5  &     4.700  &    12.0  &     0.132  &     0.973  &     0.807  &    0 \\
  1142  &  164.54578  &    1.60458  &  164.5436  &    1.5880  &    0.0394  &    21.3  &   430.1  &     3.131  &    14.0  &     0.189  &     1.118  &     0.162  &    0 \\
  1138  &  190.67690  &    2.78430  &       $-$  &       $-$  &    0.0858  &     6.9  &   308.9  &     3.400  &     9.0  &     0.325  &     0.952  &     0.707  &    0 \\
  1088  &  157.09782  &    3.75874  &  157.0948  &    3.7690  &    0.0735  &    15.4  &   591.3  &     1.798  &     5.0  &     0.218  &     1.427  &     0.494  &    1 \\
  1079  &  170.38568  &    2.88726  &  170.4047  &    2.8885  &    0.0494  &    22.6  &   576.4  &     4.866  &    12.0  &     0.135  &     1.212  &     0.492  &    0 \\
  1101  &  140.65239  &   -0.40903  &       $-$  &       $-$  &    0.0557  &    15.0  &   431.8  &     1.246  &    11.0  &     0.316  &     1.060  &     0.127  &    1 \\
  1109  &  170.72636  &    1.11440  &  170.7152  &    1.0901  &    0.0742  &    13.8  &   535.5  &     1.251  &     6.0  &     0.259  &     1.243  &     0.293  &    1 \\
  1118  &  181.11276  &    1.89600  &  181.1214  &    1.9107  &    0.0202  &    49.0  &   501.9  &    17.039  &    19.0  &     0.068  &     1.421  &     0.303  &    0 \\
  1166  &  165.18890  &   10.55318  &  165.2048  &   10.5490  &    0.0355  &    41.4  &   754.2  &     4.677  &    15.0  &     0.167  &     1.483  &     0.354  &    0 \\
  1247  &  165.23923  &   10.50548  &  165.2048  &   10.5490  &    0.0354  &    39.3  &   713.2  &     4.345  &    15.0  &     0.181  &     1.314  &     0.277  &    1 \\
  1283  &  125.74545  &    4.29911  &  125.7623  &    4.3040  &    0.0954  &    15.0  &   754.3  &     6.416  &    14.0  &     0.174  &     1.015  &     2.555  &    0 \\
  1356  &  125.84028  &    4.37247  &  125.8387  &    4.3769  &    0.0300  &    31.6  &   483.7  &     6.961  &    20.0  &     0.193  &     1.098  &     0.213  &    1 \\
  2001  &  351.08368  &   14.64713  &  351.0534  &   14.6617  &    0.0417  &    32.4  &   694.8  &    11.503  &    12.0  &     0.079  &     1.463  &     1.114  &   -1 \\
  2002  &  358.55701  &  -10.41920  &  358.5537  &  -10.4070  &    0.0762  &    20.4  &   811.8  &     8.252  &     8.0  &     0.098  &     1.298  &     2.809  &    1 \\
  2127  &  358.77844  &   -9.37558  &  358.7696  &   -9.3906  &    0.0746  &     8.3  &   323.5  &     0.720  &     5.0  &     0.317  &     1.153  &     0.135  &    0 \\
  2004  &  329.37259  &   -7.79571  &  329.3703  &   -7.7822  &    0.0579  &    20.9  &   626.6  &     3.805  &    12.0  &     0.216  &     1.174  &     0.542  &    0 \\
  2005  &   18.24821  &   15.49129  &   18.2552  &   15.5148  &    0.0433  &    22.7  &   506.5  &     4.214  &    11.0  &     0.163  &     1.264  &     0.319  &   -1 \\
  2013  &   10.46027  &   -9.30315  &   10.4438  &   -9.2373  &    0.0556  &    31.4  &   903.2  &    67.080  &    19.0  &     0.032  &     1.112  &    11.316  &    1 \\
  2015  &   29.07064  &    1.05083  &       $-$  &       $-$  &    0.0797  &    17.8  &   742.2  &     1.107  &     9.0  &     0.322  &     1.238  &     0.365  &    0 \\
  2031  &   30.57201  &   -1.12784  &   30.5779  &   -1.1174  &    0.0426  &    13.2  &   289.2  &     1.959  &     9.0  &     0.192  &     1.105  &     0.110  &   -1 \\
  2016  &   18.73999  &    0.43080  &       $-$  &       $-$  &    0.0449  &    23.9  &   552.1  &     7.170  &    15.0  &     0.114  &     1.136  &     0.544  &    0 \\
  2141  &   20.09640  &   -0.07920  &   20.0900  &   -0.0822  &    0.0779  &    10.4  &   421.6  &     0.557  &     4.0  &     0.301  &     1.306  &     0.144  &    1 \\
  2020  &  328.52530  &   -8.64287  &  328.5050  &   -8.6421  &    0.0740  &    13.1  &   504.9  &     1.689  &     6.0  &     0.247  &     1.296  &     0.403  &    0 \\
  2026  &   14.06715  &   -1.25537  &   14.1457  &   -1.2683  &    0.0444  &    40.9  &   933.0  &    36.308  &    29.0  &     0.058  &     1.091  &     3.914  &    0 \\
  2032  &    7.36845  &   -0.21261  &    7.3811  &   -0.2333  &    0.0597  &    14.8  &   458.1  &     1.035  &     9.0  &     0.315  &     1.149  &     0.136  &    0 \\
  2081  &    5.63777  &   -0.92657  &       $-$  &       $-$  &    0.0580  &     7.1  &   214.4  &     0.609  &     5.0  &     0.290  &     1.095  &     0.081  &    0 \\
  2054  &   22.88717  &    0.55597  &   22.9058  &    0.5454  &    0.0794  &    12.4  &   513.7  &     1.817  &     8.0  &     0.196  &     1.092  &     0.425  &    1 \\
  2030  &  333.69534  &   13.84087  &  333.7197  &   13.8406  &    0.0260  &    30.2  &   400.2  &     2.865  &    10.0  &     0.184  &     1.575  &     0.087  &   -1 \\
  2036  &  337.29065  &    0.07890  &  337.2749  &    0.0816  &    0.0578  &    12.7  &   381.0  &     1.230  &     6.0  &     0.307  &     1.205  &     0.147  &    1 \\
  2047  &   24.31406  &   -9.19761  &   24.3182  &   -9.1959  &    0.0406  &    21.9  &   455.4  &     6.109  &    10.0  &     0.090  &     1.218  &     0.374  &    1 \\
  2049  &  334.06497  &   -9.33325  &  334.0629  &   -9.3436  &    0.0841  &    12.6  &   555.0  &     7.479  &     6.0  &     0.113  &     1.199  &     2.222  &    1 \\
  2125  &  334.41815  &   -9.19784  &  334.4119  &   -9.2061  &    0.0945  &     3.8  &   189.4  &     0.957  &     5.0  &     0.299  &     0.952  &     0.346  &    0 \\
  2050  &   17.51319  &   13.97812  &   17.5271  &   14.0028  &    0.0588  &    24.9  &   759.2  &     1.186  &     6.0  &     0.251  &     2.050  &     0.352  &   -1 \\
  2069  &   16.84109  &   14.27322  &   16.8257  &   14.2764  &    0.0746  &    18.5  &   720.4  &     1.002  &     7.0  &     0.311  &     1.442  &     0.328  &   -1 \\
  2112  &  315.59933  &    0.25743  &  315.6010  &    0.2616  &    0.0510  &     8.0  &   211.7  &     1.107  &     5.0  &     0.210  &     1.140  &     0.119  &   -1 \\
  3280  &  133.67439  &   40.40579  &  133.6850  &   40.4256  &    0.0877  &     8.3  &   381.9  &     0.764  &     6.0  &     0.309  &     1.084  &     0.202  &    0 \\
  3074  &  225.28316  &   47.27660  &  225.2898  &   47.2864  &    0.0880  &    15.2  &   704.6  &     3.120  &     9.0  &     0.135  &     1.161  &     1.136  &    0 \\
  3325  &  230.12077  &   44.97098  &  230.1225  &   44.9753  &    0.0639  &    14.3  &   472.8  &     0.677  &     6.0  &     0.261  &     1.259  &     0.114  &    1 \\
  3002  &  255.63808  &   33.51668  &  255.6381  &   33.5032  &    0.0880  &    20.6  &   951.1  &     2.036  &     9.0  &     0.162  &     1.239  &     1.042  &    1 \\
  3012  &  255.67708  &   34.06002  &  255.5988  &   34.1199  &    0.0990  &    21.5  &  1126.7  &    19.775  &    16.0  &     0.043  &     1.076  &    13.537  &    0 \\
  3059  &  257.45279  &   34.45899  &  257.4865  &   34.4483  &    0.0858  &    11.8  &   529.9  &    10.775  &    12.0  &     0.062  &     0.996  &     2.728  &    0 \\
  3003  &  177.02458  &   54.64628  &       $-$  &       $-$  &    0.0597  &    17.3  &   534.9  &     2.385  &    11.0  &     0.211  &     1.130  &     0.322  &    0 \\
  3018  &  176.83920  &   55.73010  &  176.8489  &   55.7536  &    0.0517  &    25.6  &   683.7  &     3.823  &    14.0  &     0.160  &     1.197  &     0.467  &    0 \\
  3065  &  180.05794  &   56.25068  &  180.0793  &   56.2321  &    0.0649  &    21.5  &   723.8  &     3.809  &    13.0  &     0.175  &     1.112  &     0.723  &    1 \\
  3004  &  258.12006  &   64.06076  &  257.9521  &   64.0853  &    0.0801  &    27.6  &  1155.8  &    21.210  &    31.0  &     0.033  &     0.977  &     8.750  &    0 \\
  3060  &  259.62976  &   64.41811  &  259.6574  &   64.4133  &    0.0895  &     8.4  &   393.4  &     0.540  &    11.0  &     0.217  &     0.952  &     0.133  &   -1 \\
  3186  &  258.87518  &   64.66431  &  258.7888  &   64.6897  &    0.0795  &     9.8  &   409.2  &     0.568  &     9.0  &     0.162  &     1.020  &     0.118  &   -1 \\
  3599  &  255.49088  &   59.58125  &       $-$  &       $-$  &    0.0872  &    15.1  &   691.2  &     0.574  &     7.0  &     0.313  &     1.212  &     0.211  &    1 \\
  3005  &  239.58334  &   27.23342  &  239.5733  &   27.2463  &    0.0901  &    20.6  &   974.1  &    53.559  &    14.0  &     0.030  &     1.105  &    26.333  &    0 \\
  3096  &  152.56697  &   54.50182  &  152.5529  &   54.4864  &    0.0460  &    16.5  &   389.8  &     1.180  &     9.0  &     0.236  &     1.197  &     0.087  &   -1 \\
  3009  &  140.20340  &   40.66420  &       $-$  &       $-$  &    0.0740  &    12.4  &   480.6  &     1.003  &     7.0  &     0.303  &     1.184  &     0.216  &    0 \\
  3011  &  182.19380  &   53.33370  &  182.2094  &   53.3377  &    0.0820  &    12.2  &   524.1  &     0.990  &     5.0  &     0.253  &     1.377  &     0.314  &    0 \\
  3014  &  187.19733  &   51.26526  &  187.1927  &   51.2717  &    0.0858  &     9.1  &   408.5  &     0.643  &     4.0  &     0.263  &     1.235  &     0.191  &    1 \\
  3033  &  174.01463  &   55.07530  &       $-$  &       $-$  &    0.0564  &    15.1  &   440.6  &     1.184  &     9.0  &     0.270  &     1.157  &     0.137  &    0 \\
  3043  &  168.84946  &   54.44410  &  168.8787  &   54.4444  &    0.0698  &    17.1  &   621.4  &     2.282  &     8.0  &     0.155  &     1.208  &     0.491  &    1 \\
  3097  &  173.09666  &   55.96744  &  173.0826  &   55.9802  &    0.0514  &    13.7  &   364.4  &     3.027  &     8.0  &     0.134  &     1.168  &     0.269  &    0 \\
  3159  &  174.78577  &   55.66448  &  174.8184  &   55.6735  &    0.0611  &    13.9  &   439.7  &     0.692  &     6.0  &     0.305  &     1.244  &     0.102  &    1 \\
  3016  &  127.13190  &   30.43130  &  127.2076  &   30.4518  &    0.0498  &    32.0  &   821.5  &     7.418  &    14.0  &     0.107  &     1.329  &     1.084  &    0 \\
  3020  &  156.25665  &   47.84185  &  156.2658  &   47.8105  &    0.0629  &    18.8  &   613.0  &     1.977  &     9.0  &     0.176  &     1.269  &     0.357  &    0 \\
  3140  &  151.31216  &   53.14899  &       $-$  &       $-$  &    0.0449  &    27.6  &   637.4  &     1.120  &     8.0  &     0.261  &     1.741  &     0.141  &    0 \\
  3577  &  158.16068  &   53.15659  &       $-$  &       $-$  &    0.0637  &    12.0  &   397.4  &     0.856  &    10.0  &     0.321  &     1.043  &     0.113  &    0 \\
  3023  &  163.40237  &   54.86794  &  163.4248  &   54.9388  &    0.0719  &    14.9  &   556.4  &     3.177  &     8.0  &     0.121  &     1.205  &     0.684  &    0 \\
  3115  &  158.24541  &   56.74816  &  158.2825  &   56.7529  &    0.0451  &    18.4  &   426.5  &     0.432  &     5.0  &     0.330  &     1.588  &     0.042  &    1 \\
  3120  &  160.25429  &   58.29495  &  160.2388  &   58.2836  &    0.0732  &    11.7  &   444.8  &     1.230  &     5.0  &     0.176  &     1.342  &     0.289  &    0 \\
  3167  &  162.30042  &   57.83725  &       $-$  &       $-$  &    0.0732  &    13.8  &   525.5  &     0.735  &     4.0  &     0.268  &     1.737  &     0.233  &    0 \\
  3171  &  163.34550  &   56.31446  &  163.3528  &   56.3233  &    0.0765  &    10.5  &   418.0  &     0.496  &     5.0  &     0.300  &     1.270  &     0.120  &    0 \\
  3205  &  164.94170  &   53.80363  &  164.9784  &   53.8232  &    0.0723  &    11.3  &   426.9  &     0.635  &     5.0  &     0.260  &     1.322  &     0.142  &    0 \\
  3025  &  173.70541  &   49.07763  &       $-$  &       $-$  &    0.0330  &    39.1  &   661.2  &     6.992  &    16.0  &     0.131  &     1.274  &     0.354  &    1 \\
  3408  &  168.92094  &   48.57265  &       $-$  &       $-$  &    0.0740  &     9.4  &   362.7  &     0.796  &     9.0  &     0.319  &     1.007  &     0.133  &    1 \\
  3026  &  136.97687  &   52.79053  &  136.9746  &   52.7910  &    0.0979  &    11.3  &   583.1  &     1.121  &     7.0  &     0.306  &     1.105  &     0.431  &    1 \\
  3100  &  136.98473  &   49.59673  &  136.9897  &   49.5927  &    0.0352  &    24.1  &   434.2  &     1.888  &     7.0  &     0.187  &     1.522  &     0.107  &    1 \\
  3027  &  230.21770  &   48.66073  &  230.2238  &   48.6683  &    0.0737  &    17.4  &   669.8  &     5.709  &     8.0  &     0.093  &     1.216  &     1.455  &    1 \\
  3041  &  229.99155  &   51.31306  &  229.9925  &   51.3328  &    0.0776  &    13.7  &   555.0  &     0.484  &     6.0  &     0.322  &     1.238  &     0.125  &    1 \\
  3050  &  232.31100  &   52.86400  &  232.3134  &   52.8470  &    0.0734  &    16.6  &   635.1  &     1.973  &     9.0  &     0.141  &     1.201  &     0.475  &    0 \\
  3028  &  204.03470  &   59.20640  &  204.0723  &   59.2271  &    0.0704  &    23.8  &   872.3  &    12.147  &    13.0  &     0.057  &     1.197  &     3.470  &    0 \\
  3114  &  203.26436  &   60.11770  &  203.2779  &   60.1178  &    0.0719  &    13.4  &   502.7  &     1.659  &     9.0  &     0.164  &     1.083  &     0.310  &    1 \\
  3029  &  183.70267  &   59.90620  &  183.7677  &   59.9160  &    0.0599  &    14.1  &   436.8  &     1.303  &    10.0  &     0.251  &     1.090  &     0.161  &    0 \\
  3031  &  247.15930  &   39.55122  &  247.3164  &   39.5735  &    0.0305  &    48.4  &   754.6  &    95.497  &    29.0  &     0.020  &     1.115  &     3.984  &    1 \\
  3051  &  247.43703  &   40.81166  &  247.4344  &   40.8138  &    0.0304  &    38.5  &   596.9  &     6.874  &    21.0  &     0.106  &     1.146  &     0.247  &    1 \\
  3182  &  244.50175  &   41.39219  &  244.5108  &   41.3899  &    0.0614  &    11.0  &   351.8  &     0.795  &     5.0  &     0.210  &     1.305  &     0.114  &    0 \\
  3326  &  245.76302  &   37.92238  &  245.7615  &   37.9223  &    0.0312  &    38.1  &   608.5  &     2.444  &    10.0  &     0.142  &     1.627  &     0.133  &    1 \\
  3583  &  242.80766  &   36.97338  &  242.8176  &   36.9768  &    0.0673  &    14.6  &   512.2  &     0.855  &     5.0  &     0.263  &     1.387  &     0.180  &    1 \\
  3084  &  118.36082  &   29.35946  &  118.3380  &   29.3811  &    0.0607  &    24.8  &   781.3  &     6.563  &    10.0  &     0.100  &     1.390  &     1.443  &    0 \\
  3094  &  254.93312  &   32.61532  &  254.9353  &   32.6184  &    0.0976  &    17.0  &   874.8  &     4.491  &     8.0  &     0.088  &     1.204  &     2.549  &    1 \\
  3038  &  191.85095  &   54.98703  &       $-$  &       $-$  &    0.0833  &    14.3  &   625.3  &     1.308  &    10.0  &     0.282  &     1.097  &     0.370  &    0 \\
  3163  &  146.70914  &   43.42387  &       $-$  &       $-$  &    0.0725  &     8.3  &   314.9  &     0.490  &     5.0  &     0.331  &     1.114  &     0.084  &    1 \\
  3422  &  195.66270  &   62.49424  &  195.6442  &   62.5186  &    0.0765  &     7.8  &   310.2  &     0.592  &     6.0  &     0.280  &     1.048  &     0.106  &    1 \\
  3071  &  122.41201  &   34.92701  &  122.4113  &   34.9286  &    0.0824  &     9.2  &   398.3  &     3.746  &     7.0  &     0.144  &     1.052  &     0.859  &    1 \\
  3176  &  122.53550  &   35.27528  &  122.5413  &   35.2943  &    0.0841  &    14.2  &   626.8  &     1.310  &     5.0  &     0.287  &     1.511  &     0.523  &    0 \\
  3055  &  116.67855  &   30.99707  &  116.6630  &   30.9956  &    0.0581  &    22.5  &   675.1  &     1.182  &     5.0  &     0.272  &     2.216  &     0.336  &    0 \\
  3057  &  242.40390  &   53.04123  &  242.4066  &   53.0429  &    0.0631  &    14.3  &   467.9  &     1.608  &     9.0  &     0.175  &     1.134  &     0.237  &    0 \\
  3195  &  242.03407  &   49.20098  &  242.0678  &   49.2017  &    0.0598  &     6.7  &   206.4  &     0.429  &     4.0  &     0.289  &     1.114  &     0.064  &    1 \\
  3088  &  146.68938  &   54.42693  &  146.6944  &   54.4797  &    0.0463  &    23.3  &   555.7  &     3.576  &    10.0  &     0.152  &     1.249  &     0.319  &    1 \\
  3357  &  140.86415  &   54.82936  &  140.8650  &   54.8234  &    0.0457  &    19.7  &   462.3  &     1.121  &     9.0  &     0.278  &     1.218  &     0.088  &    1 \\
  3069  &  168.79204  &   61.11019  &       $-$  &       $-$  &    0.0554  &    15.4  &   441.3  &     0.830  &     9.0  &     0.281  &     1.166  &     0.093  &    0 \\
  3079  &  254.08789  &   39.27516  &  254.0892  &   39.2780  &    0.0622  &    13.5  &   436.0  &     1.642  &     8.0  &     0.179  &     1.160  &     0.234  &    0 \\
  3083  &  212.95599  &   52.81670  &  212.9562  &   52.8180  &    0.0760  &    11.8  &   470.1  &     0.799  &     7.0  &     0.249  &     1.119  &     0.171  &    1 \\
  3155  &  213.97035  &   50.32380  &  213.9704  &   50.3376  &    0.0745  &    12.6  &   488.6  &     0.647  &     5.0  &     0.268  &     1.400  &     0.167  &    0 \\
  3430  &  212.04359  &   52.68005  &  212.0370  &   52.6660  &    0.0819  &     5.3  &   225.9  &     0.380  &     4.0  &     0.279  &     1.052  &     0.094  &    1 \\
  3091  &  244.33394  &   34.90165  &  244.3308  &   34.9082  &    0.0309  &    32.6  &   514.8  &     5.982  &    14.0  &     0.089  &     1.341  &     0.242  &   -1 \\
  3092  &  129.04651  &   38.53475  &  129.0240  &   38.5458  &    0.0568  &    15.2  &   447.6  &     1.365  &     7.0  &     0.216  &     1.295  &     0.180  &    0 \\
  3247  &  126.50238  &   40.98111  &       $-$  &       $-$  &    0.0571  &    10.4  &   306.6  &     0.677  &     5.0  &     0.262  &     1.195  &     0.074  &    1 \\
  3103  &  243.49199  &   49.18961  &       $-$  &       $-$  &    0.0577  &    17.0  &   507.4  &     0.848  &     6.0  &     0.296  &     1.507  &     0.139  &    0 \\
  3113  &  212.51746  &   41.75580  &  212.5277  &   41.7565  &    0.0936  &    11.6  &   572.5  &     0.833  &     5.0  &     0.266  &     1.339  &     0.350  &    0 \\
  3122  &  122.59694  &   42.27387  &  122.5905  &   42.2771  &    0.0638  &    15.1  &   501.0  &     2.867  &     8.0  &     0.153  &     1.158  &     0.446  &    1 \\
  3131  &  217.08293  &   45.98601  &       $-$  &       $-$  &    0.0749  &     7.5  &   294.7  &     0.576  &     8.0  &     0.329  &     0.991  &     0.094  &    1 \\
  3271  &  215.39972  &   44.70806  &  215.4122  &   44.7125  &    0.0915  &     8.1  &   388.9  &     0.499  &     5.0  &     0.293  &     1.141  &     0.153  &    0 \\
  3389  &  217.45209  &   53.96503  &  217.4551  &   53.9706  &    0.0429  &    15.0  &   332.2  &     0.434  &     4.0  &     0.255  &     1.873  &     0.041  &    0 \\
  3143  &  261.86160  &   58.51655  &       $-$  &       $-$  &    0.0279  &    32.9  &   469.2  &     0.409  &     6.0  &     0.246  &     2.769  &     0.027  &    0 \\
  3152  &  258.84579  &   57.41119  &  258.7895  &   57.4464  &    0.0293  &    38.9  &   582.5  &    13.278  &    16.0  &     0.041  &     1.271  &     0.484  &    1 \\
  3375  &  260.85498  &   56.97455  &  260.8525  &   56.9782  &    0.0282  &    36.4  &   525.0  &     2.289  &    11.0  &     0.115  &     1.685  &     0.098  &   -1 \\
  3249  &  118.93488  &   41.20394  &  118.9279  &   41.2039  &    0.0740  &    14.0  &   541.4  &     1.115  &     6.0  &     0.263  &     1.344  &     0.282  &    0 \\
  3149  &  159.30116  &   50.12058  &  159.2658  &   50.0983  &    0.0451  &    20.1  &   466.5  &     1.549  &    10.0  &     0.216  &     1.248  &     0.122  &   -1 \\
  3157  &  138.28223  &   47.70844  &  138.2839  &   47.7101  &    0.0513  &    13.8  &   366.1  &     1.933  &     8.0  &     0.205  &     1.126  &     0.166  &    1 \\
  3166  &  238.75801  &   41.57834  &       $-$  &       $-$  &    0.0340  &    35.8  &   623.0  &     1.640  &    10.0  &     0.283  &     1.793  &     0.119  &    0 \\
  3168  &  131.08771  &   51.40593  &  131.0757  &   51.4143  &    0.0971  &     7.9  &   405.9  &     0.402  &     3.0  &     0.324  &     1.318  &     0.165  &    1 \\
  3173  &  186.12561  &   66.56688  &  186.1431  &   66.5724  &    0.0870  &     9.5  &   432.4  &     0.438  &     5.0  &     0.328  &     1.215  &     0.133  &   -1 \\
  3196  &  244.63882  &   43.08144  &  244.6446  &   43.0932  &    0.0603  &    16.4  &   511.9  &     0.508  &     4.0  &     0.333  &     2.023  &     0.123  &    0 \\
  3222  &  246.85513  &   42.67971  &  246.8527  &   42.6750  &    0.0313  &    17.0  &   271.5  &     2.122  &    12.0  &     0.137  &     1.092  &     0.064  &    0 \\
  3208  &  170.56416  &   67.22186  &  170.5571  &   67.2284  &    0.0552  &     9.3  &   264.5  &     0.778  &     5.0  &     0.192  &     1.203  &     0.085  &   -1 \\
  3212  &  117.48110  &   29.42018  &       $-$  &       $-$  &    0.0631  &    12.0  &   392.4  &     0.529  &     5.0  &     0.325  &     1.363  &     0.089  &    0 \\
  3283  &  135.32254  &   58.27975  &  135.3272  &   58.2702  &    0.0977  &    14.7  &   755.9  &     1.181  &     7.0  &     0.251  &     1.271  &     0.621  &    0 \\
  3516  &  263.05081  &   59.94155  &  263.0525  &   59.9425  &    0.0290  &     9.7  &   143.6  &     0.254  &     4.0  &     0.215  &     1.372  &     0.013  &    0 \\
  3315  &  258.62753  &   65.28972  &  258.6241  &   65.2898  &    0.0799  &     9.5  &   396.0  &     0.372  &     7.0  &     0.182  &     1.078  &     0.082  &   -1 \\
  3349  &  251.93350  &   29.94190  &       $-$  &       $-$  &    0.0996  &    16.8  &   883.6  &     4.197  &     7.0  &     0.114  &     1.363  &     2.843  &    0 \\
  3425  &  261.37708  &   53.02419  &       $-$  &       $-$  &    0.0612  &    12.6  &   398.8  &     0.636  &     8.0  &     0.235  &     1.097  &     0.080  &    1 \\
 \hline
\end{longtable}

\section{ Examples of our procedure for detecting clusters in the RASS}
In the following, we show two example images where decisions about the X-ray
detections had to be made by eye. The first example deals with the process of
determining of the X-ray centre of the cluster (Section \ref{Xcent}).  Some
X-ray sources, which are although inside the 5 arcmin aperture of the BCGs, are
judged as contaminating sources  rather than the X-ray centres  of the clusters
(Fig. B1). The second example deals with the procedure of distinguishing
contaminating X-ray sources from the extended X-ray emission of the cluster
(Fig. B2).

\begin{figure}
\includegraphics{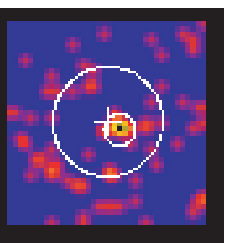}\\
\caption{The RASS image of the C4 cluster 1094. The cross shows the position of
the BCG and the big circle enclosed the region of the cluster within
a distance  $R_{200} $($\sim$ 8 arcimin) from the cluster centre.
The small circle shows the X-ray point source detected near the cluster centre.
The angular distance from the X-ray source to the position of the BCG
is $\sim$3 arcmin.  This X-ray source is judged to be a contaminant
and not  associated with
the X-ray emission of the cluster. }
\end{figure}

\begin{figure}
\includegraphics{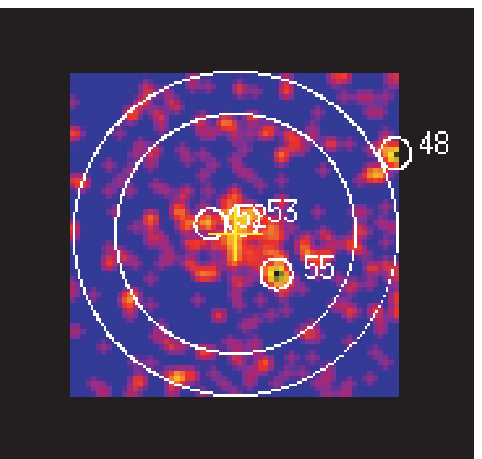}\\
\caption{The RASS image of the C4 cluster 1049. The cross shows the position of
the BCG and the small circles show the ML detected X-ray sources. The inner big
circle shows the region within a distance   $R_{200}$ ($\sim$ 16.5 arcimin)
from the cluster centre, while  the outer circle has a radius of $R_{200}+6$
arcmin. The area inside the ring from $R_{200}$ to $R_{200}+6$ arcmin is used
for the background estimation. In this image, sources 48 and 55 are considered
as contaminants,  while  sources 52 and 53 are considered as part of the
extended X-ray emission from the cluster. }
\end{figure}

\end{appendix}

\end{document}